\documentclass[10pt, journal,a4paper,final,oneside,twocolumn]{IEEEtran}
\usepackage{mathrsfs}
\usepackage{pifont}
\usepackage{bbding}
\usepackage{tipa}
\usepackage{color}
\usepackage{cite}
\usepackage{epsfig}
\usepackage{graphicx}
\usepackage{url}
\usepackage{amsfonts}
\usepackage{multicol}
\usepackage{float}
\usepackage{times}
\usepackage{cite}
\usepackage{psfrag}
\usepackage{subfigure}
\usepackage{stfloats}
\usepackage{amsmath}
\usepackage{citesort}
\usepackage{footnote}
\usepackage{array}
\usepackage{amsmath,epsfig}
\usepackage{stmaryrd}
\usepackage{amssymb}
\usepackage{epic}
\usepackage{graphicx}
\usepackage{curves}
\usepackage{balance}
\usepackage{algorithm}
\usepackage{algorithmic}

\begin{document}
\newtheorem{theorem}{Theorem}
\newtheorem{proposition}{Proposition}
\newtheorem{definition}{Definition}
\newtheorem{lemma}{Lemma}
\newtheorem{corollary}{Corollary}
\newtheorem{remark}{Remark}
\newtheorem{construction}{Construction}
\newtheorem{problem}{Problem}
\newtheorem{alg}{Algorithm}[section]

\newcommand{\supp}{\mathop{\rm supp}}
\newcommand{\sinc}{\mathop{\rm sinc}}
\newcommand{\spann}{\mathop{\rm span}}
\newcommand{\essinf}{\mathop{\rm ess\,inf}}
\newcommand{\esssup}{\mathop{\rm ess\,sup}}
\newcommand{\Lip}{\rm Lip}
\newcommand{\sign}{\mathop{\rm sign}}
\newcommand{\osc}{\mathop{\rm osc}}
\newcommand{\R}{{\mathbb{R}}}
\newcommand{\Z}{{\mathbb{Z}}}
\newcommand{\C}{{\mathbb{C}}}
\title{Pricing and Resource Allocation via Game Theory for a Small-Cell Video Caching System}
\author{{Jun~Li,~\emph{Member, IEEE},~He (Henry) Chen,~\emph{Member, IEEE},~Youjia Chen,\\~Zihuai Lin,~\emph{Senior Member, IEEE},~Branka Vucetic,~\emph{Fellow, IEEE}, and Lajos Hanzo,~\emph{Fellow, IEEE}}
\thanks{Jun~Li is with the School of Electronic and Optical Engineering, Nanjing University of Science and Technology, Nanjing, CHINA, 210094. E-mail: jun.li@njust.edu.cn.}
\thanks{He Chen, Youjia Chen, Zihuai Lin, and Branka Vucetic are with the School of Electrical and Information Engineering, The University of Sydney, AUSTRALIA. E-mail: \{he.chen, youjia.chen, zihuai.lin, branka.vucetic\}@sydney.edu.au.} \thanks{Lajos Hanzo is with the Department of Electronics and Computer Science, University of Southampton, U.K. E-mail: lh@ecs.soton.ac.uk.}
\thanks{This work is partially supported by the National Natural Science Foundation of China (No. 61501238, No. 61271230, No. 61472190), by the Jiangsu Provincial Science Foundation Project BK20150786, by the Specially Appointed Professor Program in Jiangsu Province, 2015, by the open research fund of National Key Laboratory of Electromagnetic Environment (No. 201500013), by the open research fund of National Mobile Communications Research Laboratory, Southeast University (No. 2013D02), by the Australian Research Council (No. DP120100405 and No. DP150104019), and by the Faculty of Engineering and IT Early Career Researcher Scheme 2016, The University of Sydney.}}

\markboth{IEEE Journal on Selected Areas in Communications, Accepted for Publication}
{Li \MakeLowercase{\textit{et al.}}: Pricing and Resource Allocation via Game Theory for a Small-Cell Video Caching System}
\maketitle
\begin{abstract}
Evidence indicates that downloading on-demand videos accounts for a dramatic increase in data traffic over cellular networks. Caching popular videos in the storage of small-cell base stations (SBS), namely, small-cell caching, is an efficient technology for reducing the transmission latency whilst mitigating the redundant transmissions of popular videos over back-haul channels. In this paper, we consider a commercialized small-cell caching system consisting of a network service provider (NSP), several video retailers (VR), and mobile users (MU). The NSP leases its SBSs to the VRs for the purpose of making profits, and the VRs, after storing popular videos in the rented SBSs, can provide faster local video transmissions to the MUs, thereby gaining more profits. We conceive this system within the framework of Stackelberg game by treating the SBSs as a specific type of resources. We first model the MUs and SBSs as two independent Poisson point processes, and develop, via stochastic geometry theory, the probability of the specific event that an MU obtains the video of its choice directly from the memory of an SBS. Then, based on the probability derived, we formulate a Stackelberg game to jointly maximize the average profit of both the NSP and the VRs. Also, we investigate the Stackelberg equilibrium by solving a non-convex optimization problem. With the aid of this game theoretic framework, we shed light on the relationship between four important factors: the optimal pricing of leasing an SBS, the SBSs allocation among the VRs, the storage size of the SBSs, and the popularity distribution of the VRs. Monte-Carlo simulations show that our stochastic geometry-based analytical results closely match the empirical ones. Numerical results are also provided for quantifying the proposed game-theoretic framework by showing its efficiency on pricing and resource allocation.
\end{abstract}

\section{Introduction}\label{sec:introduction}
Wireless data traffic is expected to increase exponentially in the next few years driven by a staggering proliferation of mobile users (MU) and their bandwidth-hungry mobile applications. There is evidence that streaming of on-demand videos by the MUs is the major reason for boosting the tele-traffic over cellular networks~\cite{Golrezaei:CM13}. According to the prediction of mobile data traffic by Cisco, mobile video streaming will account for $72\%$ of the overall mobile data traffic by $2019$. The on-demand video downloading involves repeated wireless transmission of videos that are requested multiple times by different users in a completely asynchronous manner, which is different from the transmission style of live video streaming.

Often, there are numerous repetitive requests of popular videos from the MUs, such as online blockbusters, leading to redundant video transmissions. The redundancy of data transmissions can be reduced by locally storing popular videos, known as caching, into the storage of intermediate network nodes, effectively forming a local caching system~\cite{Golrezaei:CM13,Xiaofei:CM14}. The local caching brings video content closer to the MUs and alleviates redundant data transmissions via redirecting the downloading requests to the intermediate nodes.

Generally, wireless data caching consists of two stages: data placement and data delivery~\cite{Ali:ALLCONF13}. In the data placement stage, popular videos are cached into local storages during off-peak periods, while during the data delivery stage, videos requested are delivered from the local caching system to the MUs. Recent works advanced the caching solutions of both device-to-device (D2D) networks and wireless sensor networks~\cite{Golrezaei:ICC12,Ji:arXiv1305,Mingyue:ISIT13}. Specifically, in~\cite{Golrezaei:ICC12} a caching scheme was proposed for a D2D based cellular network relaying on the MUs' caching of popular video content. In this scheme, the D2D cluster size was optimized for reducing the downloading delay. In~\cite{Ji:arXiv1305,Mingyue:ISIT13}, the authors proposed novel caching schemes for wireless sensor networks, where the protocol model of~\cite{Gupta:IT00} was adopted.

Since small-cell embedded architectures will dominate in future cellular networks, known as heterogeneous networks (HetNet)~\cite{Boccardi:ICM14,Damnjanovic:IWC11,Hanzo:10,Bayat:TCOM14,Mirahmadi:TCOM14,Gupta:TCOM14}, caching relying on small-cell base stations (SBS), namely, small-cell caching, constitutes a promising solution for HetNets. The advantages brought about by small-cell caching are threefold. Firstly, popular videos are placed closer to the MUs when they are cached in SBSs, hence reducing the transmission latency. Secondly, redundant transmissions over SBSs' back-haul channels, which are usually expensive~\cite{Liebsch:ICC12}, can be mitigated. Thirdly, the majority of video traffic is offloaded from macro-cell base stations to SBSs.

In~\cite{Shuan:TIT13}, a small-cell caching scheme, named `Femtocaching', is proposed for a cellular network having embedded SBSs, where the data placement at the SBSs is optimized in a centralized manner for the sake of reducing the transmission delay imposed. However,~\cite{Shuan:TIT13} considers an idealized system, where neither the interference nor the impact of wireless channels is taken into account. The associations between the MUs and the SBSs are pre-determined without considering the specific channel conditions encountered. In~\cite{Bastug:PPPCACH14}, small-cell caching is investigated in the context of stochastic networks. The average performance is quantified with the aid of stochastic geometry~\cite{Stoyan:book,Haenggi:JSAC09}, where the distribution of network nodes is modeled by Poisson point process (PPP). However, the caching strategy of~\cite{Bastug:PPPCACH14} assumes that the SBSs cache the same content, hence leading to a sub-optimal solution.

As detailed above, current research on wireless caching mainly considers the data placement issue optimized for reducing the downloading delay. However, the entire caching system design involves numerous issues apart from data placement. From a commercial perspective, it will be more interesting to consider the topics of pricing for video streaming, the rental of local storage, and so on. A commercialized caching system may consist of video retailers (VR), network service providers (NSP) and MUs. The VRs, e.g., Youtube, purchase copyrights from video producers and publish the videos on their web-sites. The NSPs are typically operators of cellular networks, who are in charge of network facilities, such as macro-cell base stations and SBSs.

In such a commercial small-cell caching system, the VRs' revenue is acquired from providing video streaming for the MUs. As the central servers of the VRs, which store the popular videos, are usually located in the backbone networks and far away from the MUs, an efficient solution is to locally cache these videos, thereby gaining more profits from providing faster local transmissions. In turn, these local caching demands raised by the VRs offer the NSPs profitable opportunities from leasing their SBSs. Additionally, the NSPs can save considerable costs due to reduced redundant video transmissions over SBSs' back-haul channels. In this sense, both the VRs and NSPs are the beneficiaries of the local caching system. However, each entity is selfish and wishes to maximize its own benefit, raising a competition and optimization problem among these entities, which can be effectively solved within the framework of game theory.

We note that game theory has been successfully applied to wireless communications for solving resource allocation problems. In~\cite{Vazquez:TWC10}, the authors propose a dynamic spectrum leasing mechanism via power control games. In~\cite{Xin:JSAC12}, a price-based power allocation scheme is proposed for spectrum sharing in Femto-cell networks based on Stackelberg game. Game theoretical power control strategies for maximizing the utility in spectrum sharing networks are studied in~\cite{Niyato:TWC08,Niyato:TMC09}.

In this paper, we propose a commercial small-cell caching system consisting of an NSP, multiple VRs and MUs. We optimize such a system within the framework of Stackelberg game by viewing the SBSs as a specific type of resources for the purpose of video caching. Generally speaking, Stackelberg game is a strategic game that consists of a leader and several followers competing with each other for certain resources~\cite{Fudenberg:gametheory}. The leader moves first and the followers move subsequently. Correspondingly, in our game theoretic caching system, we consider the NSP to be the leader and the VRs as the followers. The NSP sets the price of leasing an SBS, while the VRs compete with each other for renting a fraction of the SBSs.

To the best of the authors' knowledge, our work is the first of its kind that optimizes a caching system with the aid of game theory. Compared to many other game theory based resource allocation schemes, where the power, bandwidth and time slots are treated as the resources, our work has a totally different profit model, established based on our coverage derivations. In particular, our contributions are as follows.
\begin{enumerate}
\item{By following the stochastic geometry framework of~\cite{Stoyan:book,Haenggi:JSAC09}, we model the MUs and SBSs in the network as two different ties of a Poisson point process (PPP)~\cite{Daley:book}. Under this network model, we define the concept of a successful video downloading event when an MU obtains the requested video directly from the storage of an SBS. Then we quantify the probability of this event based on stochastic geometry theory.}
\item{Based on the probability derived, we develop a profit model of our caching system and formulate the profits gained by the NSP and the VRs from SBSs leasing and renting.}
\item{A Stackelberg game is proposed for jointly maximizing the average profit of the NSP and the VRs. Given this game theoretic framework, we investigate a non-uniform pricing scheme, where the price charged to different VRs varies.}
\item{Then we investigate the Stackelberg equilibrium of this scheme via solving a non-convex optimization problem. It is interesting to observe that the optimal solution is related both to the storage size of each SBS and to the popularity distribution of the VRs.}
\item{Furthermore, we consider an uniform pricing scheme. We find that although the uniform pricing scheme is inferior to the non-uniform one in terms of maximizing the NSP's profit, it is capable of reducing more back-haul costs compared with the latter and achieves the maximum sum profit of the NSP and the VRs.}
\end{enumerate}

The rest of this paper is organized as follows. We describe the system model in Section~\ref{sec:system_model} and establish the related profit model in Section~\ref{sec:profit_model}. We then formulate Stackelberg game for our small-cell caching system in Section~\ref{sec:problem_formulation}. In Section~\ref{sec:optimization_non_uniform}, we investigate Stackelberg equilibrium for the non-uniform pricing scheme by solving a non-convex optimization problem, while in Section~\ref{sec:further}, we further consider the uniform pricing scheme.  Our simulations and numerical results are detailed in Section~\ref{sec:numerical}, while our conclusions are provided in Section~\ref{sec:conclusion}.

\section{System Model}\label{sec:system_model}
\begin{figure}
\centering
\includegraphics[width=3.3in,angle=0]{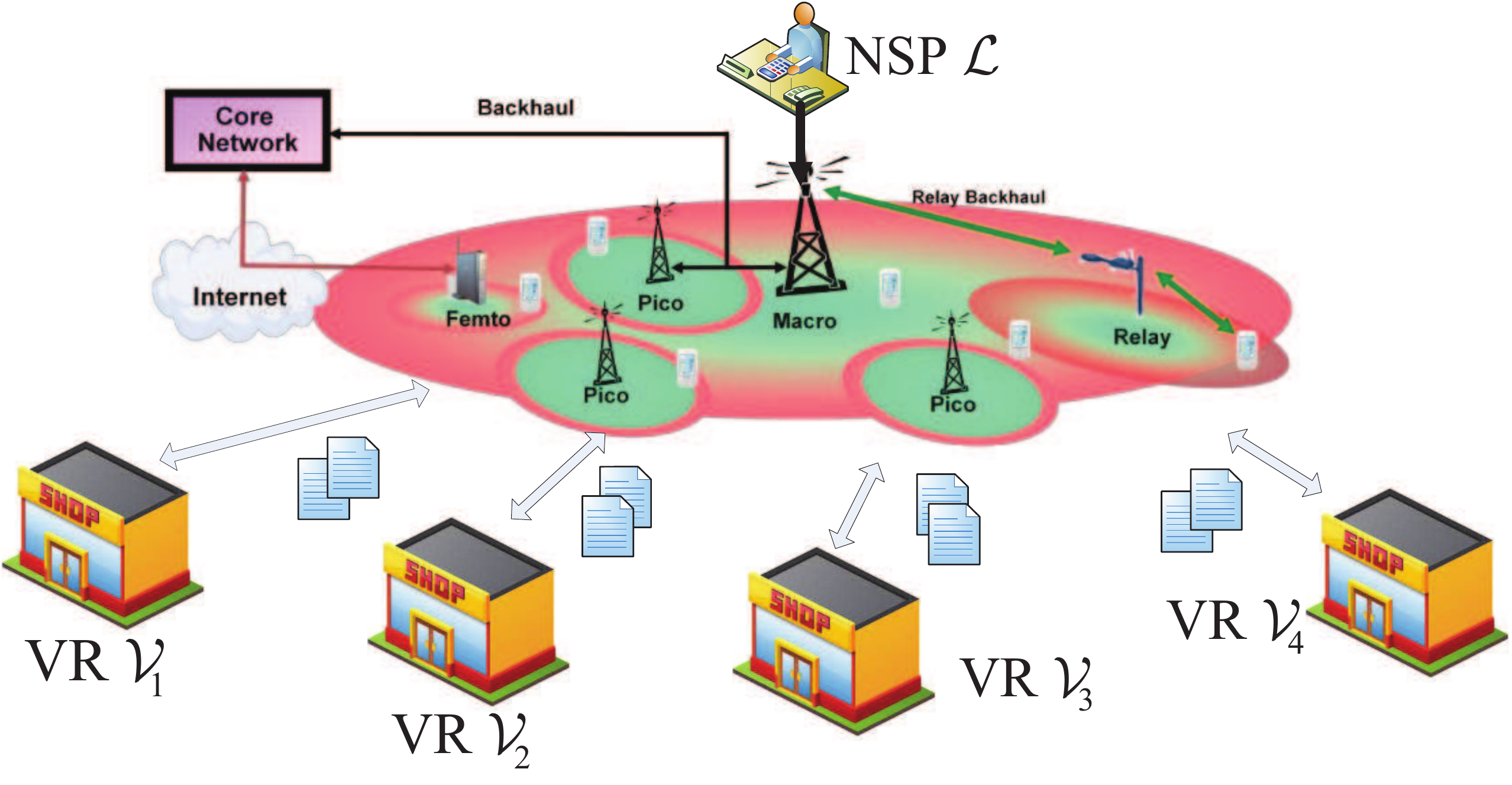}
\caption{An example of the small-cell caching system with four VRs.}\label{fig:network_M}
\end{figure}

We consider a commercial small-cell caching system consisting of an NSP, $V$ VRs, and a number of MUs. Let us denote by $\mathcal L$ the NSP, by $\boldsymbol{\mathcal V}=\{\mathcal V_1,\mathcal V_2,\cdots,\mathcal V_V\}$ the set of the VRs, and by $\mathcal M$ one of the MUs. Fig.~\ref{fig:network_M} shows an example of our caching system relying on four VRs. In such a system, the VRs wish to rent the SBSs from $\mathcal L$ for placing their videos. Both the NSP and each VR aim for maximizing their profits.

There are three stages in our system. In the first stage, the VRs purchase the copyrights of popular videos from video producers and publish them on their web-sites. In the second stage, the VRs negotiate with the NSP on the rent of SBSs for caching these popular videos. In the third stage, the MUs connect to the SBSs for downloading the desired videos. We will particulary focus our attention on the second and third stages within this game theoretic framework.
\subsection{Network Model}
Let us consider a small-cell based caching network composed of the MUs and the SBSs owned by $\mathcal L$, where each SBS is deployed with a fixed transmit power $P$ and the storage of $Q$ video files. Let us assume that the SBSs transmit over the channels that are orthogonal to those of the macro-cell base stations, and thus there is no interference incurred by the macro-cell base stations. Also, assume that these SBSs are spatially distributed according to a homogeneous PPP (HPPP) $\Phi$ of intensity $\lambda$. Here, the intensity $\lambda$ represents the number of the SBSs per unit area. Furthermore, we model the distribution of the MUs as an independent HPPP $\Psi$ of intensity $\zeta$.

The wireless down-link channels spanning from the SBSs to the MUs are independent and identically distributed (\emph{i.i.d.}), and modeled as the combination of path-loss and Rayleigh fading. Without loss of generality, we carry out our analysis for a typical MU located at the origin. The path-loss between an SBS located at $x$ and the typical MU is denoted by $\Vert x\Vert^{-\alpha}$, where $\alpha$ is the path-loss exponent. The channel power of the Rayleigh fading between them is denoted by $h_x$, where $h_x\sim\exp(1)$. The noise at an MU is Gaussian distributed with a variance $\sigma^2$.

We consider the steady-state of a saturated network, where all the SBSs keep on transmitting data in the entire frequency band allocated. This modeling approach for saturated networks characterizes the worst-case scenario of the real systems, which has been adopted by numerous studies on PPP analysis, such as~\cite{Haenggi:JSAC09}. Hence, the received signal-to-interference-plus-noise ratio (SINR) at the typical MU from an SBS located at $x$ can be expressed as
\begin{equation}\label{equ:SINR_typical_MU}
\rho(x)=\frac{P h_{x}\Vert x\Vert^{-\alpha}}{\sum_{x'\in\Phi\backslash x}Ph_{x'}\Vert x'\Vert^{-\alpha}+\sigma^2}.
\end{equation}
The typical MU is considered to be ``covered'' by an SBS located at $x$ as long as $\rho(x)$ is no lower than a pre-set SINR threshold $\delta$, i.e.,
\begin{equation}\label{equ:coverage}
\rho(x)\ge \delta.
\end{equation}
Generally, an MU can be covered by multiple SBSs. Note that the SINR threshold $\delta$ defines the highest delay of downloading a video file. Since the quality and code rate of a video clip have been specified within the video file, the download delay will be the major factor predetermining the QoS perceived by the mobile users. Therefore, we focus our attention on the coverage and SINR in the following derivations.
\subsection{Popularity and Preferences}
We now model the popularity distribution, i.e., the distribution of request probabilities, among the popular videos to be cached. Let us denote by $\boldsymbol{\mathcal F}=\{\mathcal F_1,\mathcal F_2,\cdots,\mathcal F_N\}$ the file set consisting of $N$ video files, where each video file contains an individual movie or video clip that is frequently requested by MUs. The popularity distribution of $\boldsymbol{\mathcal F}$ is represented by a vector $\textbf{t}=[t_{1},t_{2},\cdots,t_{N}]$. That is, the MUs make independent requests of the $n$-th video $\mathcal F_n$, $n=1,\cdots,N$, with the probability of $t_{n}$. Generally, $\textbf t$ can be modeled by the Zipf distribution~\cite{Cha:2007} as
\begin{equation}\label{equ:Zipf_File}
t_n=\frac{1/n^\beta}{\sum_{j=1}^{N}1/j^\beta},\quad\forall n,
\end{equation}
where the exponent $\beta$ is a positive value, characterizing the video popularity. A higher $\beta$ corresponds to a higher content reuse, where the most popular files account for the majority of download requests. From Eq.~(\ref{equ:Zipf_File}), the file with a smaller $n$ corresponds to a higher popularity.

Note that each SBS can cache at most $Q$ video files, and usually $Q$ is no higher than the number of videos in $\boldsymbol{\mathcal F}$, i.e., we have $Q\le N$. Without loss of generality, we assume that ${N}/{Q}$ is an integer. The $N$ files in $\boldsymbol{\mathcal F}$ are divided into $F={N}/{Q}$ file groups (FG), with each FG containing $Q$ video files. The $n$-th video, $\forall n\in\{(f-1)Q+1,\cdots,fQ\}$, is included in the $f$-th FG, $f=1,\cdots,F$. Denote by $\mathcal G_f$ the $f$-th FG, and by $p_{f}$ the probability of the MUs' requesting a file in $\mathcal G_f$, and we have
\begin{equation}\label{equ:fil_blk_pop}
p_{f}=\sum_{n=(f-1)Q+1}^{fQ}t_n,\quad\forall f.
\end{equation}
File caching is then carried out on the basis of FGs, where each SBS caches one of the $F$ FGs.

At the same time, the MUs have unbalanced preferences with regard to the $V$ VRs, i.e., some VRs are more popular than others. For example, the majority of the MUs may tend to access Youtube for video streaming. The preference distribution among the VRs is denoted by $\textbf{q}=[q_{1},q_{2},\cdots,q_{V}]$, where $q_v$, $v=1,\cdots,V$, represents the probability that the MUs prefer to download videos from $\mathcal V_v$. The preference distribution $\textbf q$ can also be modeled by the Zipf distribution. Hence, we have
\begin{equation}\label{equ:Zipf_VR}
q_v=\frac{1/v^\gamma}{\sum_{j=1}^{V}1/j^\gamma},\quad\forall v,
\end{equation}
where $\gamma$ is a positive value, characterizing the preference of the VRs. A higher $\gamma$ corresponds to a higher probability of accessing the most popular VRs.
\subsection{Video Placement and Download}
Next, we introduce the small-cell caching system with its detailed parameters. In the first stage, each VR purchases the $N$ popular videos in $\boldsymbol{\mathcal F}$ from the producers and publishes these videos on its web-site. In the second stage, upon obtaining these videos, the VRs negotiate with the NSP $\mathcal L$ for renting its SBSs. As $\mathcal L$ leases its SBSs to multiple VRs, we denote by $\boldsymbol \tau=[\tau_{1},\tau_{2},\cdots,\tau_{V}]$ the fraction vector, where $\tau_{v}$ represents the fraction of the SBSs that are assigned to $\mathcal V_v$, $\forall v$. We assume that the SBSs rented by each VR are uniformly distributed. Hence, the SBSs that are allocated to $\mathcal V_v$ can be modeled as a ``thinned'' HPPP $\Phi_{v}$ with intensity $\tau_{v}\lambda$.

The data placements of the second stage commence during network off-peak time after the VRs obtain access to the SBSs. During the placements, each SBS will be allocated with one of the $F$ FGs. Generally, we assume that the VRs do not have the \emph{a priori} information regarding the popularity distribution of $\boldsymbol {\mathcal F}$. This is because the popularity of videos is changing periodically, and can only be obtained statistically after these videos quit the market. It is clear that each VR may have more or less some statistical information on the popularity distribution of videos based on the MUs' downloading history. However, this information will be biased due to limited sampling. In this case, the VRs will uniformly assign the $F$ FGs to the SBSs with equal probability of $\frac1 F$ for simplicity. We are interested in investigating the uniform assignment of video files for drawing a bottom line of the system performance. As the FGs are randomly assigned, the SBSs in $\Phi_{v}$ that cache the FG $\mathcal G_f$ can be further modeled as a ``more thinned'' HPPP  $\Phi_{v,f}$ with an intensity of $\frac1 F\tau_{v}\lambda$.

In the third stage, the MUs start to download videos. When an MU $\mathcal M$ requires a video of $\mathcal G_f$ from $\mathcal V_v$, it searches the SBSs in $\Phi_{v,f}$ and tries to connect to the nearest SBS that covers $\mathcal M$. Provided that such an SBS exists, the MU $\mathcal M$ will obtain this video directly from this SBS, and we thereby define this event by $\mathcal E_{v,f}$. By contrast, if such an SBS does not exist, $\mathcal M$ will be redirected to the central servers of $\mathcal V_v$ for downloading the requested file. Since the servers of $\mathcal V_v$ are located at the backbone network, this redirection of the demand will trigger a transmission via the back-haul channels of the NSP $\mathcal L$, hence  leading to an extra cost.
\section{Profit Modeling}\label{sec:profit_model}
We now focus on modeling the profit of the NSP and the VRs obtained from the small-cell caching system. The average profit is developed based on stochastically geometrical distributions of the network nodes in terms of per unit area times unit period ($/UAP$), e.g., $/month\cdot km^2$.
\subsection{Average Profit of the NSP}
For the NSP $\mathcal L$, the revenue gained from the caching system consists of two parts: 1) the income gleaned from leasing SBSs to the VRs and 2) the cost reduction due to reduced usage of the SBSs' back-haul channels. First, the leasing income$/UAP$ of $\mathcal L$ can be calculated as
\begin{equation}\label{equ:Profit_lease_rent}
S^{RT}=\sum_{j=1}^V \tau_{j}\lambda s_{j},
\end{equation}
where ${s_{j}}$ is the price per unit period charged to $\mathcal V_j$ for renting an SBS. Then we formulate the saved cost$/UAP$ due to reduced back-haul channel transmissions. When an MU demands a video in $\mathcal G_f$ from $\mathcal V_v$, we derive the probability $\Pr(\mathcal E_{v,f})$ as follows.
\begin{theorem}\label{the:Prob_cov}
The probability of the event $\mathcal E_{v,f}$, $\forall v,f$, can be expressed as
\begin{equation}\label{equ:Prob_cov}
\Pr(\mathcal E_{v,f})=\frac{\tau_{v}}{C(\delta, \alpha)(F-\tau_{v})+\emph{A}(\delta,\alpha)\tau_{v}+\tau_{v}},
\end{equation}
where we have ${A}(\delta,\alpha)\triangleq\frac{2\delta}{\alpha-2}\; {{}_2}F_{1}\left(1,1-\frac{2}{\alpha};2-\frac{2}{\alpha};-\delta\right)$ and $C(\delta,\alpha)\triangleq\frac{2}{\alpha}\delta^{\frac{2}{\alpha}}B\left(\frac{2}{\alpha},1-\frac{2}{\alpha}\right)$. Furthermore, ${{}_2}F_{1}(\cdot)$ in the function ${A}(\delta,\alpha)$ is the hypergeometric function, while the Beta function in $C(\delta,\alpha)$ is formulated as $B(x,y)=\int^{1}_{0}t^{x-1}(1-t)^{y-1}\text{d}t$.

\emph{Proof:} Please refer to Appendix~\ref{app:prof_theo_1}.\hfill$\blacksquare$
\end{theorem}

\begin{remark}\label{rmk:theom1}
From \emph{Theorem~\ref{the:Prob_cov}}, it is interesting to observe that the probability $\Pr(\mathcal E_{v,f})$ is independent of both the transmit power $P$ and the intensity $\lambda$ of the SBSs. Furthermore, since $Q$ is inversely proportional to $F$, we can enhance $\Pr(\mathcal E_{v,f})$ by increasing the storage size $Q$.
\end{remark}

We assume that there are on average $K$ video requests from each MU within unit period, and that the average back-haul cost for a video transmission is $s^{bh}$. Based on $\Pr(\mathcal E_{v,f})$ in Eq.~(\ref{equ:Prob_cov}), we obtain the cost reduction$/UAP$ for the back-haul channels of $\mathcal L$ as
\begin{equation}\label{equ:cost_back_haul}
S^{BH}=\sum_{j_1=1}^{F}\sum_{j_2=1}^{V}p_{j_1}q_{j_2}\zeta K\Pr(\mathcal E_{j_2,j_1})s^{bh}.
\end{equation}
By combining the above two items, the overall profit$/UAP$ for $\mathcal L$ can be expressed as
\begin{equation}\label{equ:Profit_lease}
S^{NSP}=S^{RT}+S^{BH}.
\end{equation}
\subsection{Average Profit of the VRs}
Note that the MUs can download the videos either from the memories of the SBSs directly or from the servers of the VRs at backbone networks via back-haul channels. In the first case, the MUs will be levied by the VRs an extra amount of money in addition to the videos' prices because of the higher-rate local streaming, namely, local downloading surcharge (LDS). We assume that the LDS of each video is set as $s^{ld}$. Then the revenue$/UAP$ for a VR $\mathcal V_v$ gained from the LDS can be calculated as
\begin{equation}\label{equ:VR_surcharge}
S_v^{LD}=\sum_{j=1}^{F}p_{j}q_{v}\zeta K\Pr(\mathcal E_{v,j})s^{ld}.
\end{equation}
Additionally, $\mathcal V_v$ pays for renting the SBSs from $\mathcal L$. The related cost$/UAP$ can be written as
\begin{equation}\label{equ:VR_renting}
{S_{v}^{RT}}=\tau_{v}\lambda{s_{v}}.
\end{equation}
Upon combining the two items, the profit$/UAP$ for $\mathcal V_v$, $\forall v$, can be expressed as
\begin{equation}\label{equ:VR_net_Profit}
S_v^{VR}=S_v^{LD}-S_v^{RT}.
\end{equation}
\section{Problem Formulation}\label{sec:problem_formulation}
In this section, we first present the Stackelberg game formulation for our price-based SBS allocation scheme. Then the equilibrium of the proposed game is investigated.
\subsection{Stackelberg Game Formulation}
Again, Stackelberg game is a strategic game that consists of a leader and several followers competing with each other for certain resources~\cite{Fudenberg:gametheory}. The leader moves first and the followers move subsequently. In our small-cell caching system, we model the NSP $\mathcal L$ as the leader, and the $V$ VRs as the followers. The NSP imposes a price vector $\textbf{s}=[s_1,s_2,\cdots,s_V]$ for the lease of its SBSs, where $s_v$, $\forall v$, has been defined in the previous section as the price per unit period charged on $\mathcal V_v$ for renting an SBS. After the price vector $\textbf{s}$ is set, the VRs update the fraction $\tau_v$, $\forall v$, that they tend to rent from $\mathcal L$.
\subsubsection{Optimization Formulation of the Leader}
Observe from the above game model that the NSP's objective is to maximize its profit $S^{NSP}$ formulated in Eq. (\ref{equ:Profit_lease}). Note that for $\forall v$, the fraction $\tau_v$ is a function of the price $s_{v}$ under the Stackelberg game formulation. This means that the fraction of the SBSs that each VR is willing to rent depends on the specific price charged to them for renting an SBS. Consequently, the NSP has to find the optimal price vector $\textbf{s}$ for maximizing its profit. This optimization problem can be summarized as follows.
\begin{problem}\label{prb:nsp_pro}
The optimization problem of maximizing $\mathcal L$'s profit can be formulated as
\begin{equation}\label{equ:NSP_optimization}
\begin{split}
\max_{\textbf{s}\succeq\textbf{0}}~&S^{NSP}(\textbf{s},\boldsymbol{\tau}),\\
\text{s.t.}~&\sum_{{j=1}}^V \tau_j\le 1.
\end{split}
\end{equation}
\end{problem}
\subsubsection{Optimization Formulation of the Followers}
The profit gained by the VR $\mathcal V_v$ in Eq. (\ref{equ:VR_net_Profit}) can be further written as
\begin{multline}\label{equ:VR_profit_rewrite}
S_v^{VR}(\tau_v,s_v)=\sum_{j=1}^{F}p_{j}q_{v}\zeta K\Pr(\mathcal E_{v,j})s^{ld}-\tau_{v}\lambda{s_{v}}\\=
\sum_{j=1}^{F}\frac{p_jq_{v}\zeta K s^{ld}\tau_{v}}{(\emph{A}(\delta,\alpha)-C(\delta, \alpha)+1)\tau_{v}+C(\delta, \alpha)F}-\lambda{s_{v}}\tau_{v}.
\end{multline}
We can see from Eq. (\ref{equ:VR_profit_rewrite}) that once the price $s_v$ is fixed, the profit of $\mathcal V_v$ depends on $\tau_v$, i.e., the fraction of SBSs that are rented by $\mathcal V_v$. If $\mathcal V_v$ increases the fraction $\tau_v$, it will gain more revenue by levying surcharges from more MUs, while at the same time, $\mathcal V_v$ will have to pay for renting more SBSs. Therefore, $\tau_v$ has to be optimized for maximizing the profit of $\mathcal V_v$. This optimization can be formulated as follows.
\begin{problem}\label{prb:vr_pro}
The optimization problem of maximizing $\mathcal V_v$'s profit can be written as
\begin{equation}\label{equ:VR_optimization}
\max_{\tau_v\ge 0}~S_v^{VR}(\tau_v,s_v).
\end{equation}
\end{problem}

\emph{Problem \ref{prb:nsp_pro}} and \emph{Problem \ref{prb:vr_pro}} together form a Stackelberg game. The objective of this game is to find the Stackelberg Equilibrium (SE) points from which neither the leader (NSP) nor the followers (VRs) have incentives to deviate. In the following, we investigate the SE points for the proposed game.
\subsection{Stackelberg Equilibrium}
For our Stackelberg game, the SE is defined as follows.
\begin{definition}\label{def:SE}
Let $\textbf{s}^\star\triangleq[s_1^\star,s_2^\star,\cdots,s_V^\star]$ be a solution for \emph{Problem \ref{prb:nsp_pro}}, and $\tau_v^\star$ be a solution for \emph{Problem \ref{prb:vr_pro}}, $\forall v$. Define $\boldsymbol\tau^\star\triangleq[\tau_1^\star,\tau_2^\star,\cdots,\tau_V^\star]$. Then the point $(\textbf{s}^\star,\boldsymbol{\tau}^\star)$ is an SE for the proposed Stackelberg game if for any $(\textbf{s},\boldsymbol{\tau})$ with $\textbf{s}\succeq\textbf{0}$ and $\boldsymbol\tau\succeq \textbf 0$, the following conditions are satisfied:
\begin{equation}\label{equ:SE_condition}
\begin{split}
&S^{NSP}(\textbf{s}^\star,\boldsymbol{\tau}^\star)\ge S^{NSP}(\textbf{s},\boldsymbol{\tau}^\star),
\\&S_v^{VR}(s_v^\star,\tau_v^\star)\ge S_v^{VR}(s_v^\star,\tau_v),~\forall v.
\end{split}
\end{equation}
\end{definition}
Generally speaking, the SE of a Stackelberg game can be obtained by finding its perfect Nash Equilibrium (NE). In our proposed game, we can see that the VRs strictly compete in a non-cooperative fashion. Therefore, a non-cooperative subgame on controlling the fractions of rented SBSs is formulated at the VRs' side. For a non-cooperative game, the NE is defined as the operating points at which no players can improve utility by changing its strategy unilaterally. At the NSP's side, since there is only one player, the best response of the NSP is to solve \emph{Problem \ref{prb:nsp_pro}}. To achieve this, we need to first find the best response functions of the followers, based on which, we solve the best response function for the leader.

Therefore, in our game, we first solve \emph{Problem \ref{prb:vr_pro}} given a price vector $\textbf{s}$. Then with the obtained best response function $\boldsymbol{\tau}^\star$ of the VRs, we solve \emph{Problem \ref{prb:nsp_pro}} for the optimal price $\textbf{s}^\star$. In the following, we will have an in-depth investigation on this game theoretic optimization.
\section{Game Theoretic Optimization}\label{sec:optimization_non_uniform}
In this section, we will solve the optimization problem in our game under the non-uniform pricing scheme, where the NSP $\mathcal L$ charges the VRs with different prices $s_1,\cdots, s_V$ for renting an SBS. In this scheme, we first solve \emph{Problem \ref{prb:vr_pro}} at the VRs, and rewrite Eq. (\ref{equ:VR_profit_rewrite}) as
\begin{equation}\label{equ:VR_pro_rew}
S_v^{VR}(\tau_v,s_v)=\frac{\Gamma_v s^{ld}\tau_{v}}{\Theta\tau_{v}+\Lambda}-\lambda{s_{v}}\tau_{v}.0
\end{equation}
where $\Gamma_v\triangleq \sum_{j=1}^Fp_jq_{v}\zeta K$, $\Theta\triangleq \emph{A}(\delta,\alpha)-C(\delta, \alpha)+1$, and $\Lambda\triangleq C(\delta, \alpha)F$. We observe that Eq. (\ref{equ:VR_pro_rew}) is a concave function over the variable $\tau_v$. Thus, we can obtain the optimal solution by solving the Karush-Kuhn-Tucker (KKT) conditions, and we have the following lemma.
\begin{lemma}\label{lem:optimal_VR}
For a given price $s_v$, the optimal solution of \emph{Problem~\ref{prb:vr_pro}} is
\begin{equation}\label{equ:optmail_VR}
\tau_v^\star=\left(\sqrt{\frac{\Gamma_v\Lambda s^{ld}}{\Theta^2\lambda}}\sqrt{\frac1 {s_v}}-\frac{\Lambda}{\Theta}\right)^+,
\end{equation}
where $(\cdot)^+\triangleq \max(\cdot,0)$.

\emph{Proof:} The optimal solution $\tau_v^\star$ of $\mathcal V_v$ can be obtained by deriving $S_{v}^{VR}$ with respect to $\tau_{v}$ and solving $\frac{\text {d} S_{v}^{VR}}{\text{d} \tau_v} = 0$ under the constraint that $\tau_v\ge 0$.\hfill$\blacksquare$
\end{lemma}

We can see from \emph{Lemma~\ref{lem:optimal_VR}} that if the price $s_v$ is set too high, i.e., $s_v\ge\frac{\Gamma_vs^{ld}}{\Lambda\lambda}$, the VR $\mathcal V_v$ will opt out for renting any SBS from $\mathcal L$ due the high price charged. Consequently, the VR $\mathcal V_v$ will not participate in the game.

In the following derivations, we assume that the LDS on each video $s^{ld}$ is set by the VRs to be the cost of a video transmission via back-haul channels $s^{bh}$. The rational behind this assumption is as follows. Since a local downloading reduce a back-haul transmission, this saved back-haul transmission can be potentially utilized to provide extra services (equivalent to the value of $s^{bh}$) for the MUs. In addition, the MUs enjoy the benefit from faster local video transmissions. In light of this, it is reasonable to assume that the MUs are willing to accept the price $s^{bh}$ for a local video transmission.

Substituting the optimal $\tau_v^\star$ of Eq. (\ref{equ:optmail_VR}) into Eq.~(\ref{equ:Profit_lease}) and carry out some further manipulations, we arrive at
\begin{align}\label{equ:profit_lease_rw}
&S^{NSP}=\sum_{j=1}^V \lambda s_j \left(\sqrt{\frac{\Gamma_j\Lambda s^{bh}}{\Theta^2\lambda}}\sqrt{\frac1 {s_j}}-\frac{\Lambda}{\Theta}\right)^++
\nonumber\\&\qquad\qquad\qquad\frac{\sum_{i=1}^Fp_iq_j\zeta Ks^{bh}\left(\sqrt{\frac{\Gamma_j\Lambda s^{bh}}{\Theta^2\lambda}}\sqrt{\frac1 {s_j}}-\frac{\Lambda}{\Theta}\right)^+}{\Theta\left(\sqrt{\frac{\Gamma_j\Lambda s^{bh}}{\Theta^2\lambda}}\sqrt{\frac1 {s_j}}-\frac{\Lambda}{\Theta}\right)^++\Lambda}\nonumber\\&=\sum_{j=1}^V \frac{\xi_i}{\Theta}\left(-{\Lambda\lambda}s_j+\left(\sqrt{s^{bh}}-\frac{s^{bh}}{\sqrt{s^{bh}}}\right)\sqrt{{\Gamma_j\Lambda\lambda s_j}}+{\Gamma_j}s^{bh}\right)\nonumber\\&=\sum_{j=1}^V \frac{\xi_i}{\Theta}\left(-{\Lambda\lambda}s_j+{\Gamma_j}s^{bh}\right),
\end{align}
where $\xi_j$ is the indicator function, with $\xi_j=1$ if $s_j<\frac{\Gamma_js^{bh}}{\Lambda\lambda}$ and $\xi_j=0$ otherwise. Upon defining the binary vector $\boldsymbol \xi\triangleq[\xi_1,\xi_2,\cdots,\xi_V]$, we can rewrite \emph{Problem \ref{prb:nsp_pro}} as follows.
\begin{problem}\label{prb:nsp_pro_rw}
Given the optimal solutions $\tau_v^\star$, $\forall v$, gleaned from the followers, we can rewrite \emph{Problem \ref{prb:nsp_pro}} as
\begin{equation}\label{equ:NSP_optimization_rw}
\begin{split}
\min_{\boldsymbol \xi,~\textbf{s}\succeq\textbf{0}}~&\sum_{j=1}^V {\xi_j}\left({\Lambda\lambda}s_j-
{\Gamma_j}s^{bh}\right),\\
\text{s.t.}~&\sum_{{j=1}}^V \xi_j\left(\sqrt{\frac{\Gamma_j\Lambda s^{bh}}{\lambda s_j}}-{\Lambda}\right)\le \Theta.
\end{split}
\end{equation}
\end{problem}

Observe from Eq.~(\ref{equ:NSP_optimization_rw}) that \emph{Problem~\ref{prb:nsp_pro_rw}} is non-convex due to $\boldsymbol \xi$. However, for a given $\boldsymbol \xi$, this problem can be solved by satisfying the KKT conditions. In the following, we commence with the assumption that $\boldsymbol \xi=\textbf{1}$, i.e., $\xi_v=1$, $\forall v$, and then we extend this result to the general case.
\subsection{Special Case: $\xi_v=1$, $\forall v$}
In this case, all the VRs are participating in the game, and we have the following optimization problem.
\begin{problem}\label{prb:nsp_pro_rw_rw}
Assuming $\xi_v=1$, $\forall v$, we rewrite \emph{Problem \ref{prb:nsp_pro_rw}} as
\begin{equation}\label{equ:NSP_optimization_rw_1}
\begin{split}
\min_{\textbf{s}\succeq\textbf{0}}~&\sum_{j=1}^V s_j,\\
\text{s.t.}~&\sum_{{j=1}}^V \sqrt{\frac{\Gamma_j}{s_j}}\le (V{\Lambda}+ \Theta)\sqrt{\frac{\lambda}{\Lambda s^{bh}}}.
\end{split}
\end{equation}
\end{problem}

The optimal solution of \emph{Problem~\ref{prb:nsp_pro_rw_rw}} is derived and given in the following lemma.
\begin{lemma}\label{lem:Prob_cov_xi_one}
The optimal solution to \emph{Problem~\ref{prb:nsp_pro_rw_rw}} can be derived as $\hat{\textbf{s}}\triangleq[\hat s_1,\cdots,\hat s_V]$, where
\begin{equation}\label{equ:lem_s_v}
\hat s_v=\frac{\Lambda s^{bh}\left(\sum_{j=1}^V\sqrt[3]{\Gamma_j}\right)^2\sqrt[3]{\Gamma_v}}{{\lambda(V\Lambda+\Theta)^2}}, \forall v.
\end{equation}

\emph{Proof:} Please refer to Appendix~\ref{app:prof_lem_2}.\hfill$\blacksquare$
\end{lemma}

Note that the solution given in \emph{Lemma~\ref{lem:Prob_cov_xi_one}} is found under the assumption that $\xi_v=1$, $\forall v$. That is, $\hat s_v$ given in Eq.~(\ref{equ:lem_s_v}) should ensure that $\tau_v^\star>0$, $\forall v$, in Eq.~(\ref{equ:optmail_VR}), i.e.,
\begin{equation}\label{equ:condition}
\frac{\Lambda s^{bh}\left(\sum_{j=1}^V\sqrt[3]{\Gamma_j}\right)^2\sqrt[3]{\Gamma_v}}{{\lambda(V\Lambda+\Theta)^2}}<\frac{\Gamma_vs^{bh}}{\Lambda\lambda}.
\end{equation}
Given the definitions of $\Gamma_v$, $\Lambda$, and $\Theta$, it is interesting to find that the inequality (\ref{equ:condition}) can be finally converted to a constraint on the storage size $Q$ of each SBS, which is formulated as
\begin{equation}\label{equ:F_constraint}
Q>\max\left\{\frac{NC(\delta,\alpha)\left(\sum_{j=1}^V\sqrt[3]{\frac{q_j}{q_v}}-V\right)}{A(\delta,\alpha)-C(\delta,\alpha)+1},~\forall v\right\}.
\end{equation}
The constraint imposed on $Q$ can be expressed in a concise manner in the following theorem.
\begin{theorem}\label{the:optimal_all_one_speical}
To make sure that $\hat s_v$ in Eq.~(\ref{equ:lem_s_v}) does become the optimal solution of \emph{Problem~\ref{prb:nsp_pro_rw_rw}} when $\xi_v=1$, $\forall v$, the sufficient and necessary condition to be satisfied is
\begin{equation}\label{equ:F_constraint_1}
Q>Q_{min}\triangleq\frac{NC(\delta,\alpha)\left(\sum_{j=1}^V\sqrt[3]{\frac{q_j}{q_V}}-V\right)}{A(\delta,\alpha)-C(\delta,\alpha)+1},
\end{equation}
where $q_V$ is the minimum value in $\textbf{q}$ according to Eq.~(\ref{equ:Zipf_VR}).

\emph{Proof:} Please refer to Appendix~\ref{app:prof_theo_2}.\hfill$\blacksquare$
\end{theorem}

\begin{remark}\label{rmk:minimum_Q}
Observe from Eq.~(\ref{equ:F_constraint_1}) that since ${\frac{q_j}{q_V}}$ increases exponentially with $\gamma$ according to Eq.~(\ref{equ:Zipf_VR}), the value of $Q_{min}$ ensuring $\xi_v=1$, $\forall v$, will increase exponentially with $\gamma/3$.
\end{remark}

Note that we have $Q\le N$. In the case that $Q_{min}$ in Eq.~(\ref{equ:F_constraint_1}) is larger than $N$ for a high VR popularity exponent $\gamma$, some VRs with the least popularity will be excluded from the game.
\subsection{Further Discussion on $Q$}
We define a series of variables $U_v$, $\forall v$, as follows:
\begin{equation}\label{equ:U_v}
U_v\triangleq\frac{NC(\delta,\alpha)\left(\sum_{j=1}^v\sqrt[3]{\frac{q_j}{q_v}}-v\right)}{A(\delta,\alpha)-C(\delta,\alpha)+1},
\end{equation}
and formulate the following lemma.
\begin{lemma}\label{lem:U_compare}
$U_v$ is a strictly monotonically-increasing function of $v$, i.e., we have $U_V>U_{V-1}>\cdots>U_1$.

\emph{Proof:} Please refer to Appendix~\ref{app:prof_lem_3}.\hfill$\blacksquare$
\end{lemma}

For the special case of the previous subsection, the optimal solution for $\xi_v=1$, $\forall v$, is found under the condition that the storage size obeys $Q>U_V$. In other words, $Q$ should be large enough such that every VR can participate in the game. However, when $Q$ reduces, some VRs have to leave the game as a result of the increased competition. Then we have the following lemma.
\begin{lemma}\label{lem:U_compare_xi_result}
When $U_{v}<Q\le U_{v+1}$, the NSP can only retain at most the $v$ VRs of $\mathcal V_1,\mathcal V_2,\cdots,\mathcal V_{v}$ in the game for achieving its optimal solution.

\emph{Proof:} Please refer to Appendix~\ref{app:prof_lem_4}.\hfill$\blacksquare$
\end{lemma}

From \emph{Lemma~\ref{lem:U_compare_xi_result}}, when we have $U_{v}<Q\le U_{v+1}$, and given that there are $u$ VRs, $u\le v$, in the game, we can have an optimal solution for $\textbf{s}$.
\begin{problem}\label{prb:nsp_pro_rw_rw_5}
when $U_{v}<Q\le U_{v+1}$ is satisfied, and given that there are $u$, $u\le v$, VRs in the game, we can formulate the following optimization problem as
\begin{equation}\label{equ:NSP_optimization_rw_rw_5}
\begin{split}
\min_{\textbf{s}\succeq\textbf{0}}~&\sum_{j=1}^u s_j,\\
\text{s.t.}~&\sum_{{j=1}}^u \sqrt{\frac{\Gamma_j}{s_j}}\le (u{\Lambda}+ \Theta)\sqrt{\frac{\lambda}{\Lambda s^{bh}}}.
\end{split}
\end{equation}
\end{problem}
Similar to the solution of \emph{Problem~\ref{prb:nsp_pro_rw_rw}}, we arrive at the optimal solution for the above problem as $\hat {\textbf s}_u\triangleq[\hat s_{1,u},\cdots,\hat s_{i,u},\cdots,\hat s_{V,u}]$, where
\begin{equation}\label{equ:lem_s_v_i}
\hat s_{i,u}=\begin{cases}\frac{\Lambda s^{bh}\left(\sum_{j=1}^u\sqrt[3]{\Gamma_j}\right)^2\sqrt[3]{\Gamma_i}}{{\lambda(u\Lambda+\Theta)^2}},&\quad i=1,\cdots,u,\\\qquad\qquad \infty,&\quad i=u+1,\cdots,V.
\end{cases}
\end{equation}
\subsection{General Case}
Let us now focus our attention on the general solution of the original optimization problem, i.e., of \emph{Problem~\ref{prb:nsp_pro_rw}}. Without loss of generality, we consider the case of $U_{v}<Q\le U_{v+1}$. Then \emph{Problem~\ref{prb:nsp_pro_rw}} is equivalent to the following problem.
\begin{problem}\label{prb:nsp_pro_rw_general}
When $U_{v}<Q\le U_{v+1}$, there are at most $v$ VRs in the game. Then \emph{Problem~\ref{prb:nsp_pro_rw}} can be converted to
\begin{equation}\label{equ:NSP_optimization_rw_general}
\begin{split}
\min_{\boldsymbol \xi,~\textbf{s}\succeq\textbf{0}}~&\sum_{j=1}^{v} {\xi_j}\left({\Lambda\lambda}s_j-
{\Gamma_j}s^{bh}\right),\\
\text{s.t.}~&\sum_{{j=1}}^{v} \xi_j\left(\sqrt{\frac{\Gamma_j\Lambda s^{bh}}{\lambda s_j}}-{\Lambda}\right)\le \Theta.
\end{split}
\end{equation}
\end{problem}

The problem in Eq.~(\ref{equ:NSP_optimization_rw_general}) is again non-convex due to the uncertainty of $\xi_u$, $u=1,\cdots, v$. We have to consider the cases, where there are $u$, $\forall u$, most popular VRs in the game. We observe that for a given $u$, \emph{Problem~\ref{prb:nsp_pro_rw_general}} converts to \emph{Problem~\ref{prb:nsp_pro_rw_rw_5}}. Therefore, to solve \emph{Problem~\ref{prb:nsp_pro_rw_general}}, we first solve \emph{Problem~\ref{prb:nsp_pro_rw_rw_5}} with a given $u$ and obtain $\hat {\textbf s}_u$ according to Eq.~(\ref{equ:lem_s_v_i}). Then we choose the optimal solution, denoted by ${\textbf{s}}^\star_{v}$, among $\hat {\textbf s}_1,\cdots,\hat {\textbf s}_v$ as the solution to \emph{Problem~\ref{prb:nsp_pro_rw_general}}, which is formulated as
\begin{multline}\label{equ:optimal_ori}
\textbf{s}_{v}^\star=\\\arg\min_{\hat {\textbf s}_u}\left\{\min\left(\sum_{j=1}^u\left({\Lambda\lambda}s_j-
{\Gamma_j}s^{bh}\right)\right),~u=1,\cdots,v\right\}.
\end{multline}

Based on the above discussions, we can see that the optimal solution $\textbf{s}^\star$ of \emph{Problem~\ref{prb:nsp_pro_rw}} is a piece-wise function of $Q$, i.e., $\textbf{s}^\star=\textbf{s}_{v}^\star$ when $U_{v}<Q\le U_{v+1}$. Now, we formulate the solution $\textbf{s}^\star=[s_1^\star,\cdots,s_V^\star]$ to \emph{Problem~\ref{prb:nsp_pro_rw}} in a general manner as follows.
\begin{equation}\label{equ:general_solution}
s_v^\star=\begin{cases}\frac{\Lambda s^{bh}\left(\sum_{j=1}^{\hat u}\sqrt[3]{\Gamma_j}\right)^2\sqrt[3]{\Gamma_v}}{{\lambda(\hat u\Lambda+\Theta)^2}},&\quad v=1,\cdots,\hat u,\\\qquad\qquad \infty,&\quad v=\hat u+1,\cdots,V,
\end{cases}
\end{equation}
where regarding $\hat u$, we have
\begin{equation}\label{equ:general_solution_Khat}
\hat u=\arg\min_u~\{S_u:u=1,2,\cdots,T\},
\end{equation}
with $S_u$ formulated as
\begin{align}\label{equ:general_solution_S_k}
S_u&=\sum_{j_1=1}^u\left(\frac{\Lambda^2 s^{bh}\left(\sum_{j_2=1}^{u}\sqrt[3]{\Gamma_{j_2}}\right)^2\sqrt[3]{\Gamma_{j_1}}}{{(u\Lambda+\Theta)^2}}-{\Gamma_{j_1}}s^{bh}\right),\nonumber\\
T&=\begin{cases}1,\quad U_1<Q\le U_2,\\\cdots,\\v,\quad U_v<Q\le U_{v+1},\\\cdots,\\V, \quad U_V<Q.
\end{cases}
\end{align}
To gain a better understanding of the optimal solution in Eq.~(\ref{equ:general_solution}), we propose a centralized algorithm at $\mathcal L$ in Table~\ref{fig:alg1} for obtaining $\textbf {s}^\star$.
\begin{remark}\label{rmk:SE}
The optimal solution $\textbf{s}^\star$ in Eq.~(\ref{equ:general_solution}), combined with the solution of $\boldsymbol \tau^\star$ given by Eq.~(\ref{equ:optmail_VR}) in \emph{Lemma~\ref{lem:optimal_VR}}, constitutes the SE for the Stackelberg game.
\end{remark}
\begin{table}[!tbh]
\begin{center}
\begin{algorithm}[H] \label{Alg:1}
\textbf{Input:}$\quad$ Storage size $Q$, number of videos $N$, VRs' preference distribution $\textbf{q}$, channel exponent $\alpha$, and pre-set threshold $\delta$.\\
\textbf{Output:}$\quad$ Optimal pricing vector $\textbf{s}^\star$.\\
\textbf{Steps:}
\begin{algorithmic}[1]
\STATE Based on $N$, $\textbf{q}$, $\alpha$, and $\delta$, the NSP calculates $U_v$, $\forall v$, according to Eq.~(\ref{equ:U_v});
\STATE By comparing $Q$ to $U_v$, the NSP obtains the value of the integer $T$ in Eq.~(\ref{equ:general_solution_S_k});
\STATE Calculate $S_u$, $u=1,2,\cdots,T$, according to Eq.~(\ref{equ:general_solution_S_k});
\STATE Compare among $S_1,\cdots,S_T$ for finding the index $\hat u$ of the minimum $S_{\hat u}$;
\STATE Based on $\hat u$, $N$, $\textbf{q}$, $\alpha$, and $\delta$, the NSP obtains the optimal solution $\textbf{s}^\star$ according to Eq.~(\ref{equ:general_solution}).
\end{algorithmic}
\caption{:}
\end{algorithm}
\end{center}
\caption{The centralized algorithm at the NSP for obtaining the optimal solution $\textbf{s}^\star$.}\label{fig:alg1}
\end{table}

Furthermore, by substituting the optimal $\textbf{s}^\star$ into the expression of $S^{NSP}$ in Eq.~(\ref{equ:profit_lease_rw}), we get
\begin{multline}\label{equ:profit_overall}
S^{NSP}(\textbf{s}^\star,\boldsymbol{\tau}^\star)=\\\frac1{\Theta}\sum_{j_1=1}^{\hat u}\left({\Gamma_{j_1}}s^{bh}-\frac{\Lambda^2 s^{bh}\left(\sum_{j_2=1}^{\hat u}\sqrt[3]{\Gamma_{j_2}}\right)^2\sqrt[3]{\Gamma_{j_1}}}{{(\hat u\Lambda+\Theta)^2}}\right).
\end{multline}
\begin{remark}\label{rmk:snsp_increase}
Since we have $\Gamma_v\propto q_{v}$, $\forall v$, and $q_v$ increases exponentially with the VR preference parameter $\gamma$ according to Eq.~(\ref{equ:Zipf_VR}), $S^{NSP}(\textbf{s}^\star,\boldsymbol{\tau}^\star)$ also increases exponentially with $\gamma$.
\end{remark}
\section{Discussions of Other Schemes}\label{sec:further}
Let us now consider two other schemes, namely, an uniform pricing scheme and a global optimization scheme.
\subsection{Uniform Pricing Scheme}
In contrast to the non-uniform pricing scheme of the previous section, the uniform pricing scheme deliberately imposes the same price on the VRs in the game. We denote the fixed price by $s$. In this case, similar to \emph{Lemma~\ref{lem:optimal_VR}}, \emph{Problem \ref{prb:vr_pro}} can be solved by
\begin{equation}\label{equ:optmail_VR_uniform}
\tau_v^\star=\left(\sqrt{\frac{\Gamma_v\Lambda s^{bh}}{\Theta^2\lambda}}\sqrt{\frac1 {s}}-\frac{\Lambda}{\Theta}\right)^+.
\end{equation}

We first focus our attention on the special case of $\xi_v=1$, $\forall v$. Then \emph{Problem~\ref{prb:nsp_pro_rw_rw}} can be converted to that of minimizing $s$ subject to the constraint $\sum_{{j=1}}^V \sqrt{\frac{\Gamma_j}{s}}\le (V{\Lambda}+ \Theta)\sqrt{\frac{\lambda}{\Lambda s^{bh}}}$. We then obtain the optimal $\hat s$ for this special case as
\begin{equation}\label{equ:uniform_s}
\hat s=\frac{\Lambda s^{bh}\left(\sum_{j=1}^V\sqrt{\Gamma_j}\right)^2}{{\lambda(V\Lambda+\Theta)^2}}.
\end{equation}
To guarantee that all the VRs are capable of participating in the game, i.e., $\xi_v=1$, $\forall v$, with the optimal price $\hat s$, we let $\hat s<\frac{\Gamma_vs^{bh}}{\Lambda\lambda}$. Then we have the following constraint on the storage $Q$ as
\begin{equation}\label{equ:F_constraint_uniform}
Q>Q'_{min}\triangleq\frac{NC(\delta,\alpha)\left(\sum_{j=1}^V\sqrt{\frac{q_j}{q_V}}-V\right)}{A(\delta,\alpha)-C(\delta,\alpha)+1}.
\end{equation}
We can see that the we require a larger storage size $Q$ in Eq.~(\ref{equ:F_constraint_uniform}) than that in Eq.~(\ref{equ:F_constraint_1}) under the non-uniform pricing scheme to accommodate all the VRs, since we have $\sum_{j=1}^V\sqrt{\frac{q_j}{q_V}}>\sum_{j=1}^V\sqrt[3]{\frac{q_j}{q_V}}$. Following \emph{Remark~\ref{rmk:minimum_Q}}, we conclude that $Q'_{min}$ of the uniform pricing scheme will increase exponentially with $\gamma/2$.

Then based on this special case, the optimal $\textbf{s}^\star=[s_1^\star,\cdots,s_V^\star]$ in the uniform pricing scheme can be readily obtained by following a similar method to that in the previous section. That is,
\begin{equation}\label{equ:general_solution_uniform}
s_v^\star=\begin{cases}\frac{\Lambda s^{bh}\left(\sum_{j=1}^{\hat u}\sqrt{\Gamma_j}\right)^2}{{\lambda(\hat u\Lambda+\Theta)^2}},&\quad v=1,\cdots,\hat u,\\\:\:\:\quad\quad \infty,&\quad v=\hat u+1,\cdots,V,
\end{cases}
\end{equation}
where regarding $\hat u$, we have
\begin{equation}\label{equ:general_solution_Khat_uniform}
\hat u=\arg\min_u~\{S_u:u=1,2,\cdots,T\},
\end{equation}
with
\begin{align}\label{equ:general_solution_S_k_uniform}
S_u&=\frac{u\Lambda^2 s^{bh}\left(\sum_{j=1}^{u}\sqrt{\Gamma_{j}}\right)^2}{{(u\Lambda+\Theta)^2}}-\sum_{j=1}^u{\Gamma_{j}}s^{bh},\nonumber\\
T&=\begin{cases}1,\quad \bar U_1<Q\le \bar U_2,\\\cdots,\\v,\quad \bar U_v<Q\le \bar U_{v+1},\\\cdots,\\V, \quad \bar U_V<Q.
\end{cases}
\end{align}
Note that $\bar U_v$ in Eq.~(\ref{equ:general_solution_S_k_uniform}) is defined as
\begin{equation}\label{equ:defin_bar_U}
\bar U_v\triangleq\frac{NC(\delta,\alpha)\left(\sum_{j=1}^v\sqrt{\frac{q_j}{q_v}}-v\right)}{A(\delta,\alpha)-C(\delta,\alpha)+1}.
\end{equation}

It is clear that the uniform pricing scheme is inferior to the non-uniform pricing scheme in terms of maximizing $S^{NSP}$. However, we will show in the following problem that the uniform pricing scheme offers the optimal solution to maximizing the back-haul cost reduction $S^{BH}$ at the NSP in conjunction with $\tau_v^\star$, $\forall v$, from the followers.
\begin{problem}\label{prb:nsp_pro_rw_uniform}
With the aid of the optimal solutions $\tau_v^\star$, $\forall v$, from the followers, the maximization on $S^{BH}$ is achieved by solving the following problem:
\begin{equation}\label{equ:NSP_optimization_rw_uniform}
\begin{split}
\min_{\boldsymbol \xi,~\textbf{s}\succeq\textbf{0}}~&\sum_{j=1}^V {\xi_j}\left(\sqrt{s^{bh}}\sqrt{{\Gamma_j\Lambda\lambda}}\sqrt{s_j}-
{\Gamma_j}s^{bh}\right),\\
\text{s.t.}~&\sum_{{j=1}}^V \xi_j\left(\sqrt{\frac{\Gamma_j\Lambda s^{bh}}{\lambda s_j}}-{\Lambda}\right)\le \Theta.
\end{split}
\end{equation}
\end{problem}

The optimal solution to \emph{Problem~\ref{prb:nsp_pro_rw_uniform}} can be readily shown to be $\textbf{s}^\star$ given in Eq.~(\ref{equ:general_solution_uniform}). This proof follows the similar procedure of the optimization method presented in the previous section. Thus it is skipped for brevity. In this sense, the uniform pricing scheme is superior to the non-uniform scheme in terms of reducing more cost on back-haul channel transmissions.
\subsection{Global Optimization Scheme}
In the global optimization scheme, we are interested in the sum profit of the NSP and VRs, which can be expressed as
\begin{equation}\label{equ:globe_optim}
\begin{split}
&S^{GLB}=S^{NSP}+\sum_{j=1}^V S^{VR}_j\\
&=\sum_{j_1=1}^{V}\sum_{j_2=1}^{F}\frac{2p_{j_2}q_{j_1}\zeta K s^{bh}\tau_{j_1}}{(\emph{A}(\delta,\alpha)-C(\delta, \alpha)+1)\tau_{j_1}+C(\delta, \alpha)F}\\&=2S^{BH}.
\end{split}
\end{equation}
Observe from Eq.~(\ref{equ:globe_optim}), we can see that the sum profit $S^{GLB}$ is twice the back-haul cost reduction $S^{BH}$, where the vector $\boldsymbol\tau$ is the only variable of this maximization problem.
\begin{problem}\label{prb:Global_opt}
The optimization of the sum profit $S^{GLB}$ can be formulated as
\begin{equation}\label{equ:Global_optimization}
\begin{split}
\max_{\boldsymbol \tau\succeq\textbf{0}}~&\sum_{j_1=1}^{V}\frac{\tau_{j_1}\sum_{j_2=1}^{F}p_{j_2}q_{j_1}\zeta K s^{bh}}{(\emph{A}(\delta,\alpha)-C(\delta, \alpha)+1)\tau_{j_1}+C(\delta, \alpha)F},\\
\text{s.t.}~&\sum_{{j=1}}^{V} \tau_j\le 1.
\end{split}
\end{equation}
\end{problem}

\emph{Problem~\ref{prb:Global_opt}} is a typical water-filling optimization problem. By relying on the classic Lagrangian multiplier, we arrive at the optimal solution as
\begin{equation}\label{equ:solution_for_water_filling}
\hat \tau_v=\left(\frac{\frac{\sqrt{q_v}}{\eta}-C(\delta, \alpha)F}{\emph{A}(\delta,\alpha)-C(\delta, \alpha)+1}\right)^+,~\forall v,
\end{equation}
where we have $\eta=\frac{\sum_{j=1}^{\bar v}{\sqrt{q_j}}}{\bar vC(\delta, \alpha)F+\emph{A}(\delta,\alpha)-C(\delta, \alpha)+1}$, and $\bar v$ satisfies the constraint of $\hat \tau_v>0$.
\subsection{Comparisons}
Let us now compare the optimal SBS allocation variable $\tau_v$ in the context of the above two schemes. First, we investigate $\tau_v^\star$ in the uniform pricing scheme. By substituting Eq.~(\ref{equ:general_solution_uniform}) into Eq.~(\ref{equ:optmail_VR_uniform}), we have
\begin{equation}\label{equ:optmail_VR_uniform_rw}
\begin{split}
\tau_v^\star&=\left(\sqrt{\frac{\Gamma_v\Lambda s^{bh}}{\Theta^2\lambda}}\sqrt{\frac1 {s_v^\star}}-\frac{\Lambda}{\Theta}\right)^+\\
&=\begin{cases}\frac{\frac{\sqrt{q_v}}{\eta'}-C(\delta, \alpha)F}{\emph{A}(\delta,\alpha)-C(\delta, \alpha)+1},&\quad v=1,\cdots,\hat u\\\qquad\quad 0,&\quad v=\hat u+1,\cdots,V,
\end{cases}
\end{split}
\end{equation}
where $\eta'=\frac{\sum_{j=1}^{\hat u}{\sqrt{q_j}}}{\hat uC(\delta, \alpha)F+\emph{A}(\delta,\alpha)-C(\delta, \alpha)+1}$, and $\hat u$ ensures $\tau_v^\star>0$.

Then, comparing $\tau_v^\star$ given in Eq.~(\ref{equ:optmail_VR_uniform_rw}) to the optimal solution $\hat\tau$ of the global optimization scheme given by Eq.~(\ref{equ:solution_for_water_filling}), we can see that these two solutions are the same. In other words, the uniform pricing scheme in fact represents the global optimization scheme in terms of maximizing the sum profit $S^{GLB}$ and maximizing the back-haul cost reduction $S^{BH}$.
\section{Numerical Results}\label{sec:numerical}
In this section, we provide both numerical as well as Monte-Carlo simulation results for evaluating the performance of the proposed schemes. The physical layer parameters of our simulations, such as the path-loss exponent $\alpha$, transmit power $P$ of the SBSs and the noise power $\sigma^2$ are similar to those of the 3GPP standards. The unit of noise power and transmit power is Watt, while the SBS and MU intensities are expressed in terms of the numbers of the nodes per square kilometer.

Explicitly, we set the path-loss exponent to $\alpha=4$, the SBS transmit power to $P=2$ Watt, the noise power to $\sigma^2=10^{-10}$ Watt, and the pre-set SINR threshold to $\delta=0.01$. For the file caching system, we set the number of files in $\boldsymbol{\mathcal F}$ to $N=500$ and set the number of VRs to $V=15$. For the network deployments, we set the intensity of the MUs to $\zeta=50/km^2$, and investigate three cases of the SBS deployments as $\lambda=10/km^2,20/km^2$ and $30/km^2$.

For the pricing system, the profit$/UAP$ is considered to be the profit gained per month within an area of one square kilometer, i.e., $/month\cdot km^2$. We note that the profits gained by the NSP and by the VRs are proportional to the cost $s^{bh}$ of back-haul channels for transmitting a video. Hence, without loss of generality, we set $s^{bh}=1$ for simplicity. Additionally, we set $K=10/month$, which is the average number of video requests from an MU per month.

We first verify our derivation of $\Pr(\mathcal E_{v,f})$ by comparing the analytical results of \emph{Theorem~\ref{the:Prob_cov}} to the Monte-Carlo simulation results. Upon verifying $\Pr(\mathcal E_{v,f})$, we will investigate the optimization results within the framework of the proposed Stackelberg game by providing numerical results.
\begin{figure}
\centering
\includegraphics[width=3.7in,angle=0]{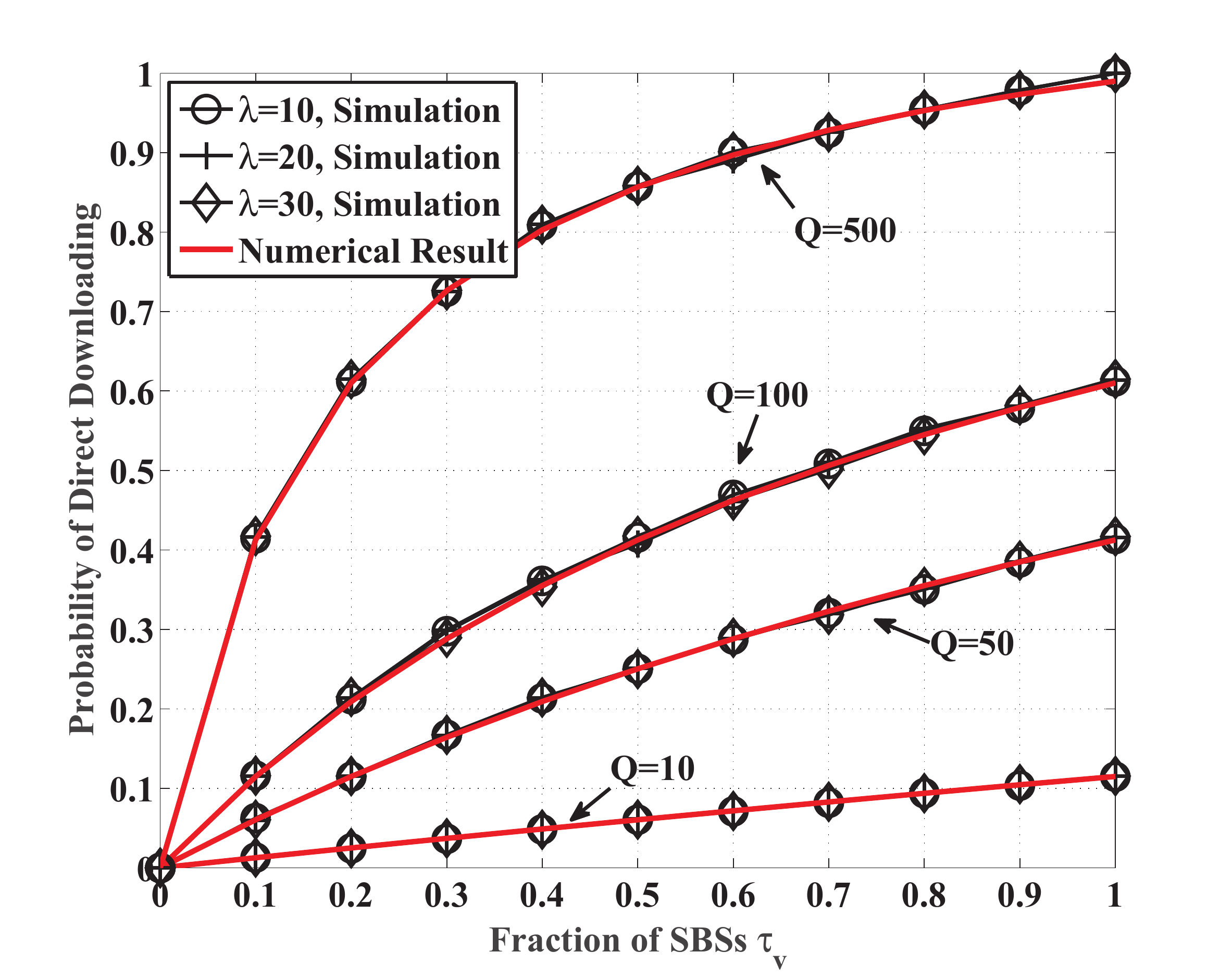}
\caption{Comparisons between the simulations and analytical results on $\Pr(\mathcal E_{v,f})$. We consider four kinds of storage size $Q$ in each SBS, i.e., $Q=10,50,100,500$, and three kinds of SBS intensity, i.e., $\lambda=10,20,30$.}\label{fig:verificiation_of_Prob}
\end{figure}
\begin{figure}
\centering
\includegraphics[width=3.7in,angle=0]{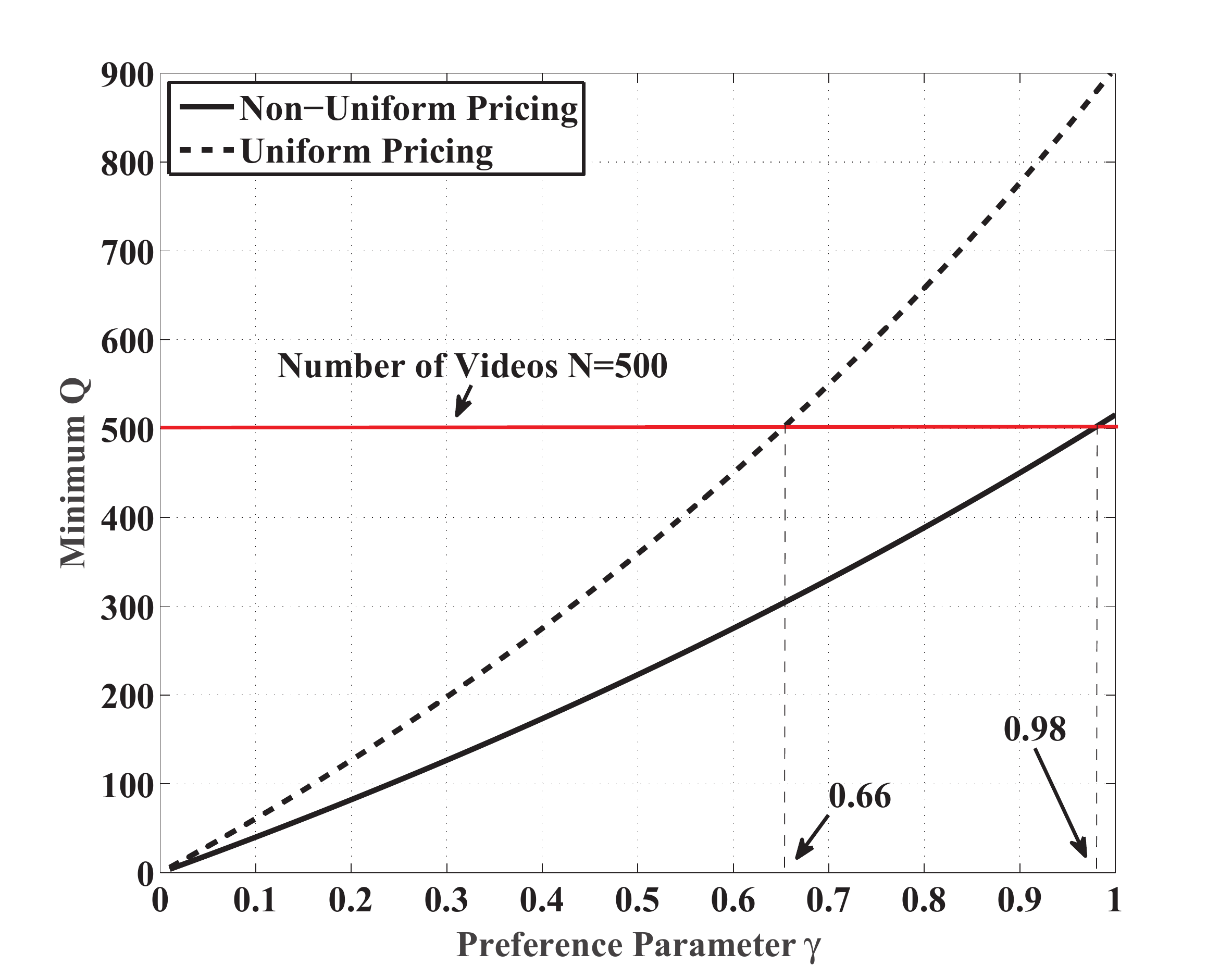}
\caption{The minimum number of $Q$ that allows all the VRs to participate in the game under different preference parameter $\gamma$. In the case that the minimum $Q$ is larger than $N$, it means that some VRs will be inevitable excluded from the game.}\label{fig:minimum_Q}
\end{figure}
\begin{figure}
\centering
\includegraphics[width=3.7in,angle=0]{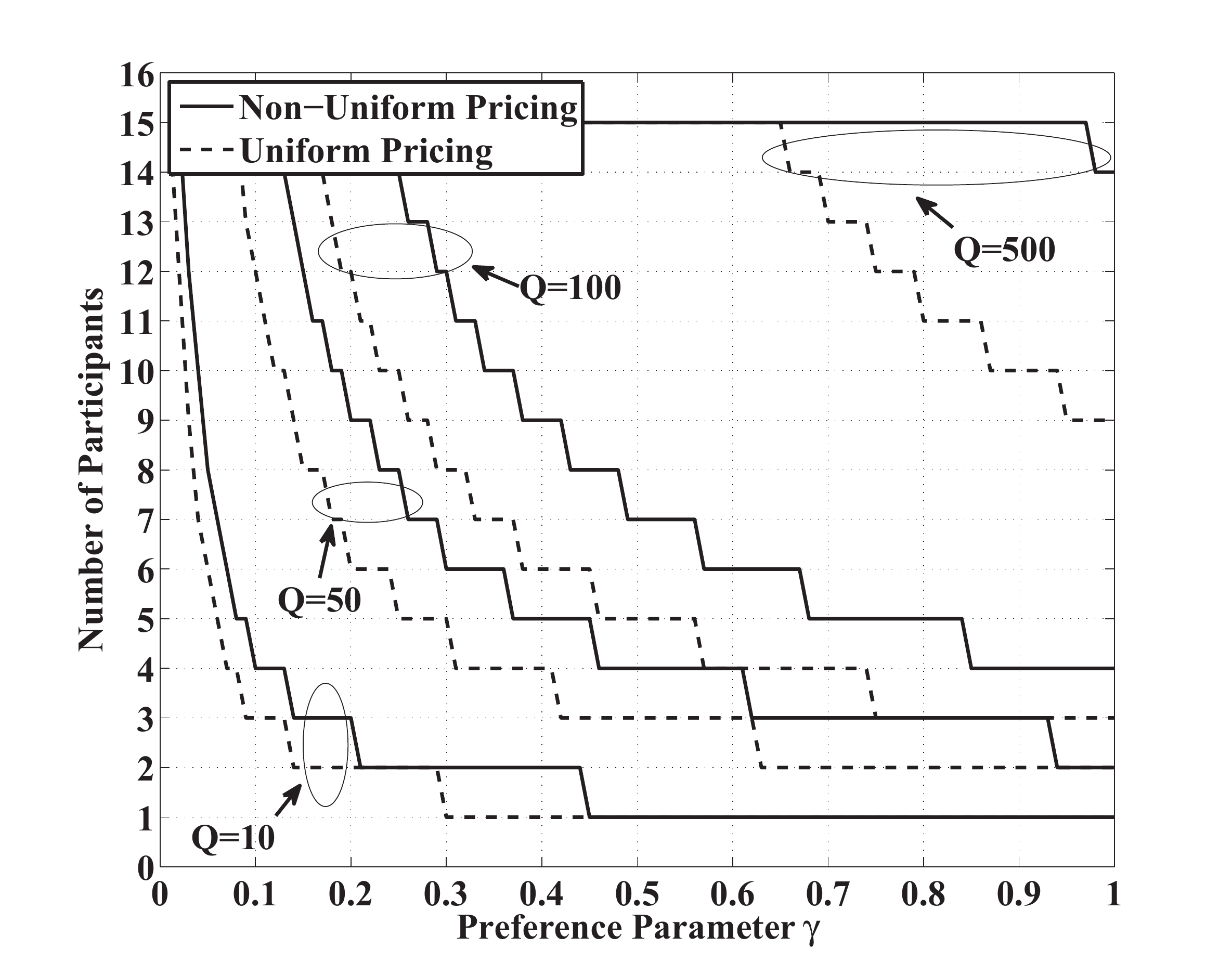}
\caption{Number of participants, i.e., the VRs that are in the game, vs. the preference parameter $\gamma$, under the two schemes. We also consider four different values of the storage size $Q$, i.e., $10,50,100,500$.}\label{fig:num_particpants}
\end{figure}
\begin{figure}
\centering
\includegraphics[width=3.7in,angle=0]{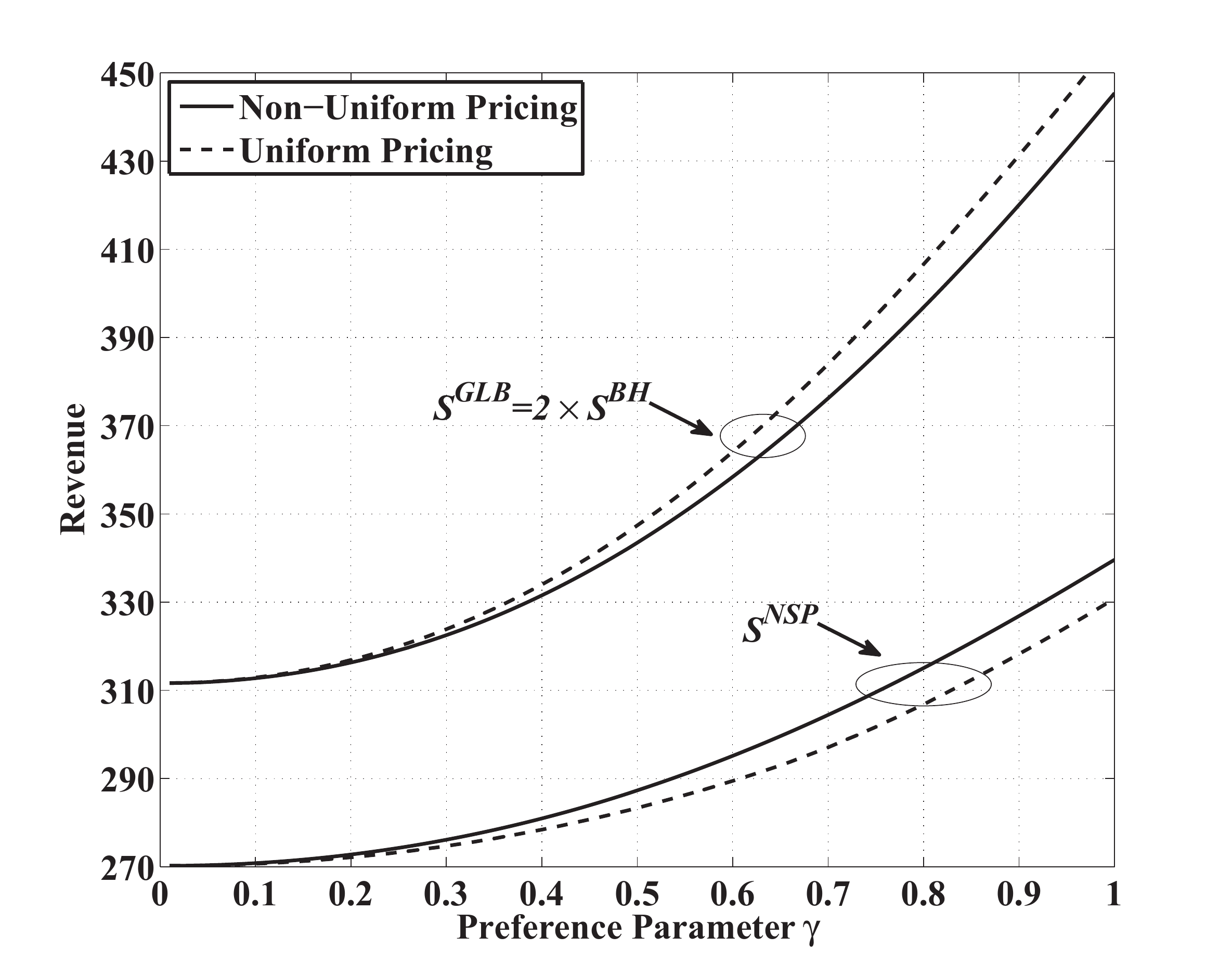}
\caption{Various revenues, including $S^{NSP}$ and $S^{GLB}$, vs. the preference parameter $\gamma$, under the two schemes.}\label{fig:profit_vs_gamma}
\end{figure}
\begin{figure}
\centering
\includegraphics[width=3.7in,angle=0]{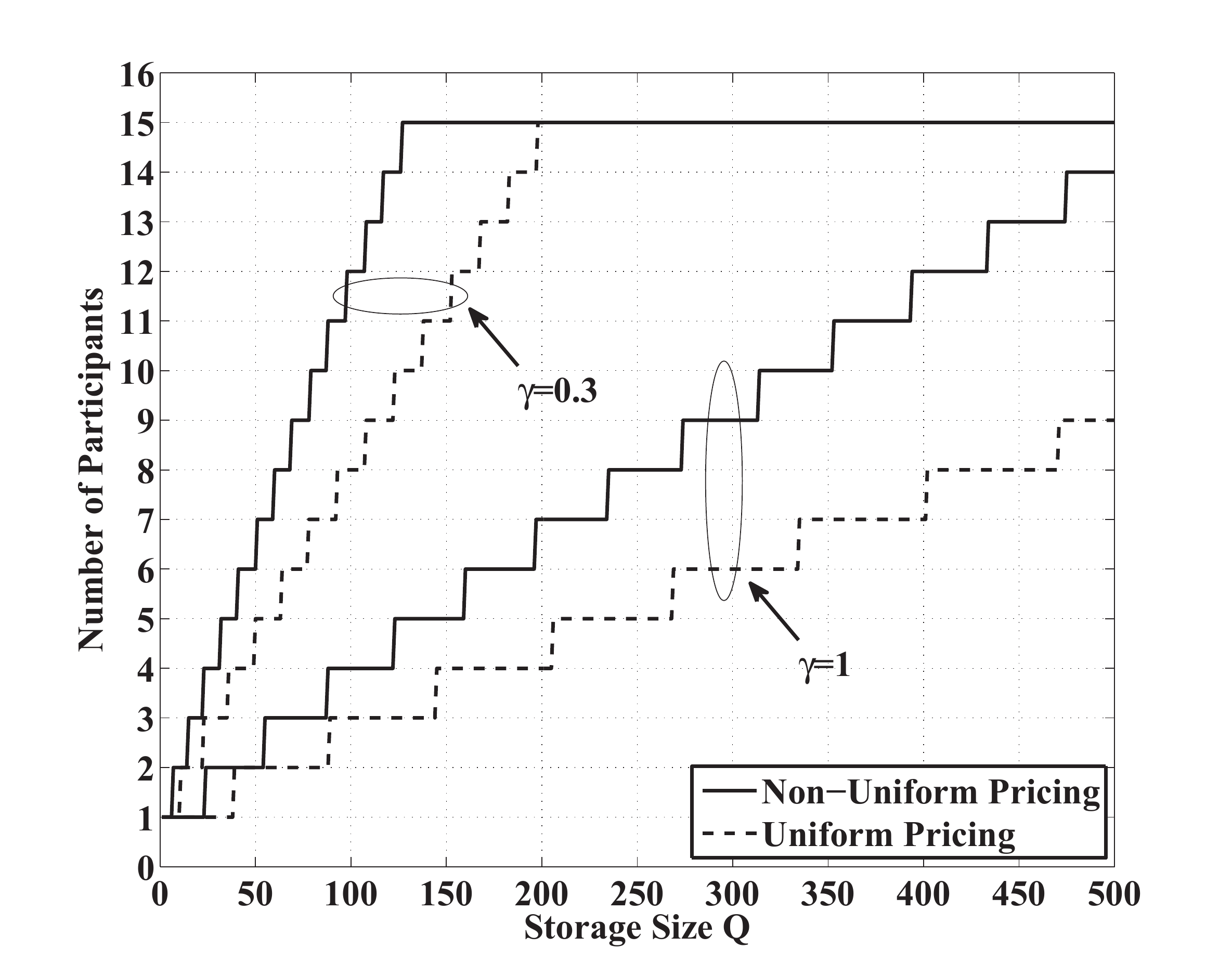}
\caption{Number of participants vs. the storage size $Q$, under the two schemes. We also consider two different values of $\gamma$, i.e., $\gamma=0.3,1$.}\label{fig:num_particpants_Q}
\end{figure}
\begin{figure}
\centering
\includegraphics[width=3.7in,angle=0]{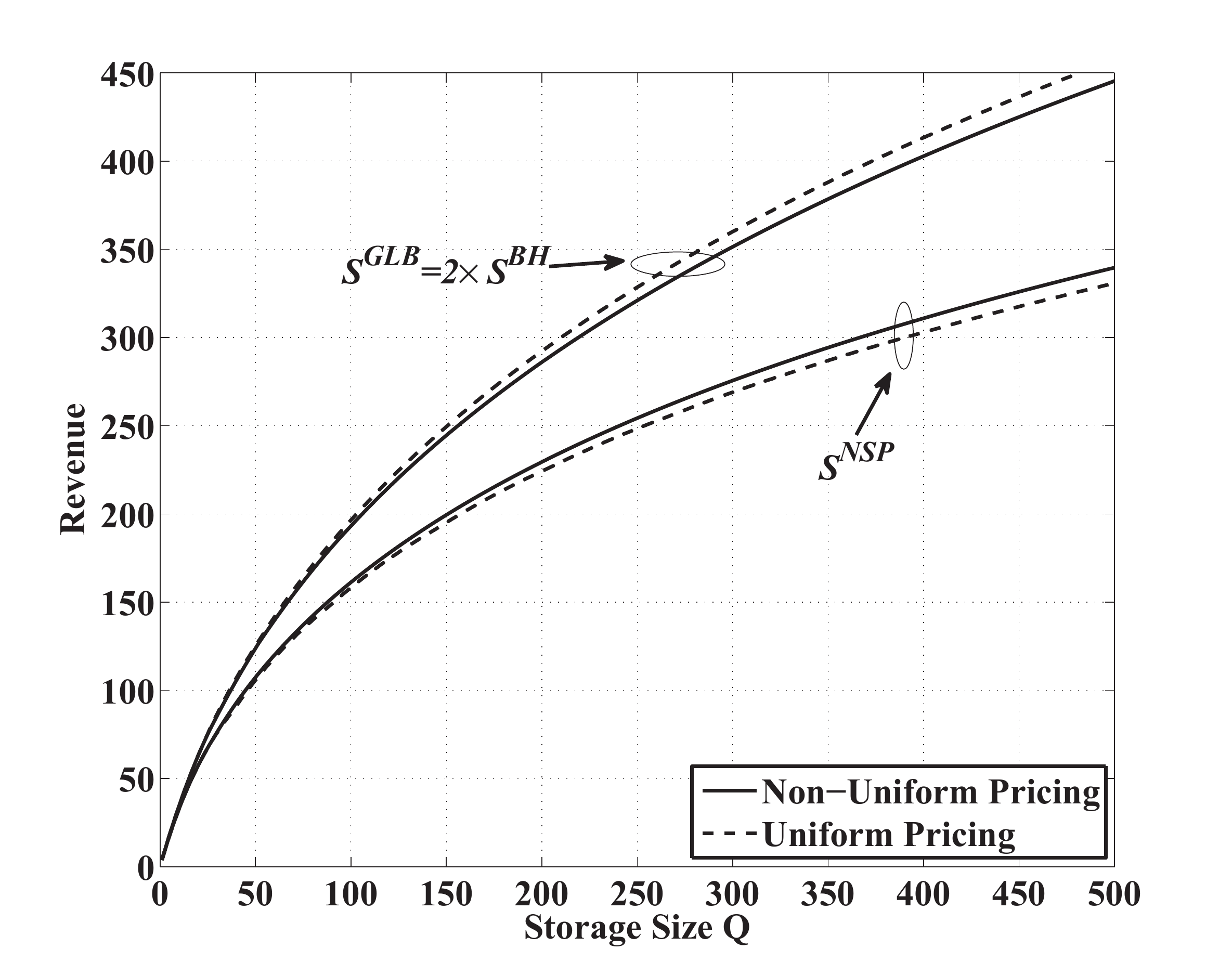}
\caption{Various revenues, including $S^{NSP}$ and $S^{GLB}$, vs. the storage size $Q$, under the two schemes.}\label{fig:profit_vs_Q}
\end{figure}
\subsection{Performance Evaluation on $\Pr(\mathcal E_{v,f})$}
For the Monte-Carlo simulations of this subsection, all the average performances are evaluated over a thousand network scenarios, where the distributions of the SBSs and the MUs change from case to case according the PPPs characterized by $\Phi$ and $\Psi$, respectively.

Note that $\Pr(\mathcal E_{v,f})$ in \emph{Theorem~\ref{the:Prob_cov}} is the probability that an MU can obtain its requested video directly from the memory of an SBS rented by $\mathcal V_v$. We can see from the expression of $\Pr(\mathcal E_{v,f})$ in Eq.~(\ref{equ:Prob_cov}) that it is a function of the fraction $\tau_v$ of the SBSs that are rented by $\mathcal V_v$. Although $\tau_v$ should be optimized according to the price charged by the NSP, here we investigate a variety of $\tau_v$ values, varying from $0$ to $1$, to verify the derivation of $\Pr(\mathcal E_{v,f})$.

Fig.~\ref{fig:verificiation_of_Prob} shows our comparisons between the simulations and analytical results on $\Pr(\mathcal E_{v,f})$. We consider four different storage sizes $Q$ in each SBS by setting $Q=10,50,100,500$. Correspondingly, we have four values for the number of file groups, i.e., $F=50,10,5,1$. Furthermore, we consider the SBS intensities of $\lambda=10,20,30$. From Fig.~\ref{fig:verificiation_of_Prob}, we can see that the simulations results closely match the analytical results derived in \emph{Theorem~\ref{the:Prob_cov}}. Our simulations show that the intensity $\lambda$ does not affect $\Pr(\mathcal E_{v,f})$, which is consistent with our analytical results. Furthermore, a larger $Q$ leads to a higher value of $\Pr(\mathcal E_{v,f})$. Hence, enlarging the storage size is helpful for achieving a higher probability of direct downloading.
\subsection{Impact of the VR Preference Parameter $\gamma$}
The preference distribution $\textbf{q}$ of the VRs defined in Eq.~(\ref{equ:Zipf_VR}) is an important factor in predetermining the system performance. Indeed, we can see from Eq.~(\ref{equ:Zipf_VR}) that this distribution depends on the parameter $\gamma$. Generally, we have $0<\gamma\le 1$, with a larger $\gamma$ representing a more uneven popularity among the VRs. First, we find the minimum $Q$ that can keep all the VRs in the game. This minimum $Q$ for the non-uniform pricing scheme (NUPS) is given by Eq.~(\ref{equ:F_constraint_1}), while the minimum $Q$ for the uniform-pricing scheme (UPS) is given by Eq.~(\ref{equ:F_constraint_uniform}). From the two equations, this minimum $Q$ increases exponentially with $\gamma/3$ in the NUPS, while it also increases exponentially with a higher exponent of $\gamma/2$ in the UPS. Fig.~\ref{fig:minimum_Q} shows this minimum $Q$ for different values of the VR preference parameter $\gamma$.

We can see that the UPS needs a larger $Q$ than the NUPS for keeping all the VRs. This gap increases rapidly with the growth of $\gamma$. For example, for $\gamma=0.3$, the uniform pricing scheme requires almost $80$ more storages, while for $\gamma=0.6$, it needs $200$ more. We can also observe in Fig.~\ref{fig:minimum_Q} that for $\gamma>0.66$ in the UPS and for $\gamma>0.98$ in the NUPS, the minimum $Q$ becomes larger than the overall number of videos $N$. In both cases, since we have $Q\le N$ ($Q>N$ results in the same performance as $Q=N$), some unpopular VRs will be excluded from the game.

Next, we study the number of VR participants that stay in the game for the two schemes upon increasing $\gamma$. We can see from Fig.~\ref{fig:num_particpants} that the number of VR participants keeps going down upon increasing $\gamma$ in the both schemes. The NUPS always keeps more VRs in the game than the UPS under the same $\gamma$. At the same time, by considering $Q=10,50,100,500$, it is shown that for a given $\gamma$, a higher $Q$ will keep more VRs in the game.

Fig.~\ref{fig:profit_vs_gamma} shows two kinds of revenues gained by the two schemes for a given storage of $Q=500$, namely, the global profit $S^{GLB}$ defined in Eq.~(\ref{equ:globe_optim}) and the profit of the NSP $S^{NSP}$ defined in Eq.~(\ref{equ:Profit_lease}). Recall that we have $S^{GLB}=2S^{BH}$ according to Eq.~(\ref{equ:globe_optim}). We can see that the revenues of both schemes increase exponentially upon increasing $\gamma$, as stated in \emph{Remark~\ref{rmk:snsp_increase}}. As our analytical result shows, the profit $S^{NSP}$ gained by the NUPS is optimal and thus it is higher than that gained by the UPS, while the UPS maximizes both $S^{GLB}$ and $S^{BH}$. Fig.~\ref{fig:profit_vs_gamma} verifies the accuracy of our derivations.
\begin{figure}
\centering
\includegraphics[width=3.7in,angle=0]{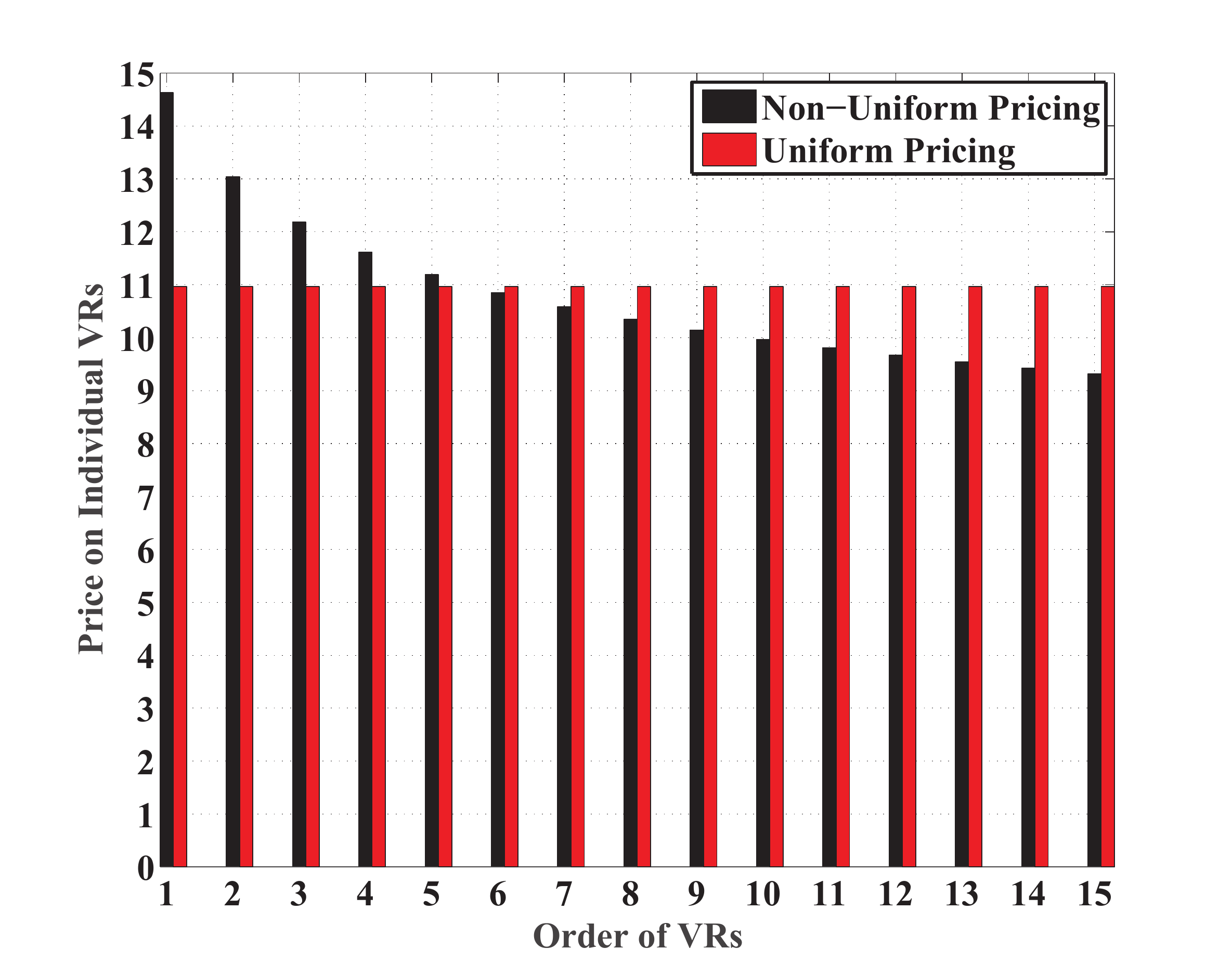}
\caption{Price charged on each VR for renting an SBS per month.}\label{fig:individual_price}
\end{figure}
\begin{figure}
\centering
\includegraphics[width=3.7in,angle=0]{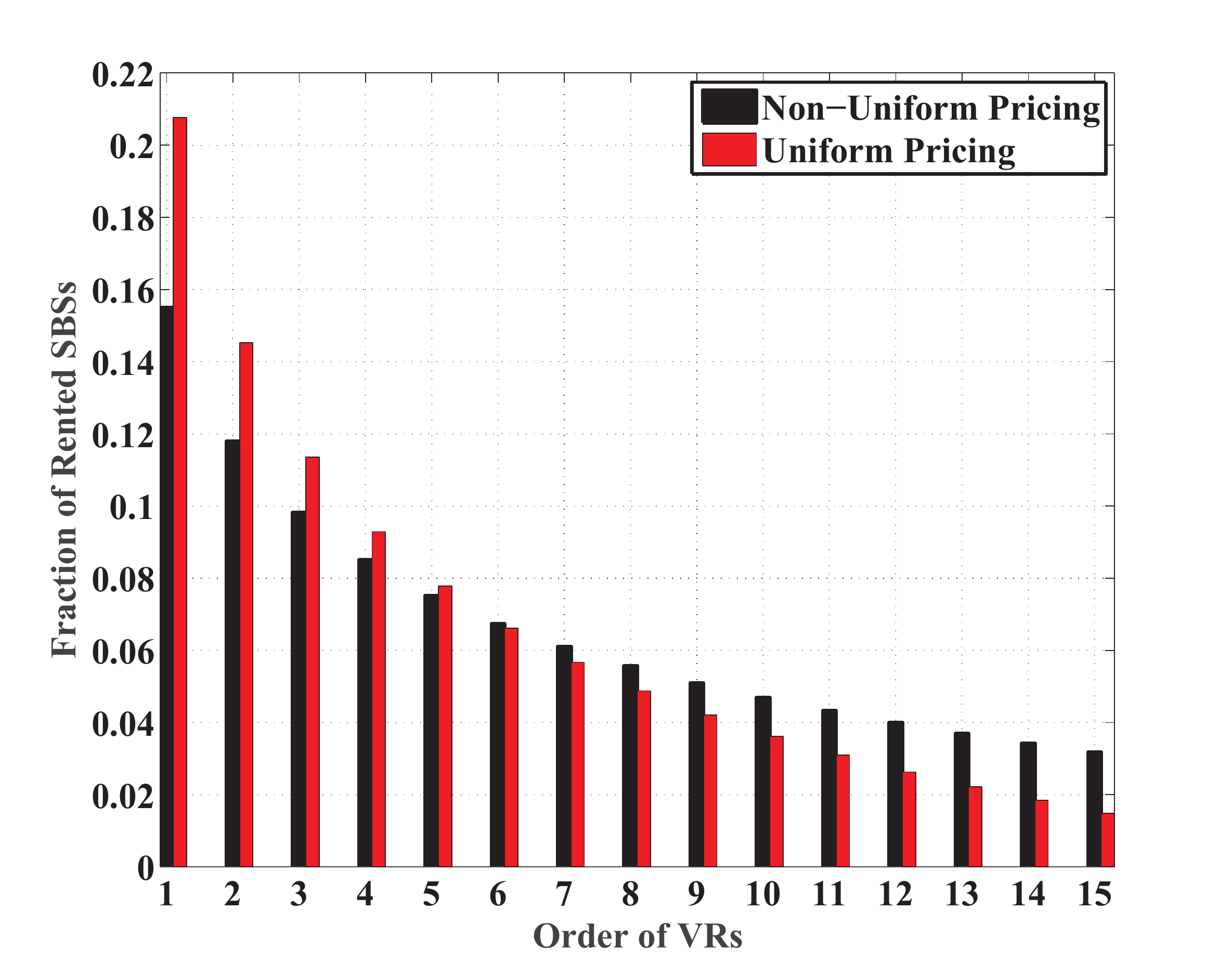}
\caption{The fraction of SBSs that are rented by each VR.}\label{fig:individual_fraction}
\end{figure}
\subsection{Impact of the Storage Size $Q$}
Since $\gamma$ is a network parameter that is relatively fixed, the NSP can adapt the storage size $Q$ for controlling its performance. In this subsection, we investigate the performance as a function of $Q$. Fig.~\ref{fig:num_particpants_Q} shows the number of participants in the game versus $Q$, where $\gamma=0.3$ and $1$ are considered. It is shown that for a larger $Q$, more VRs are able to participate in the game. Again, the NUPS outperforms the UPS owing to its capability of accommodating more VRs for a given $Q$. By comparing the scenarios of $\gamma=0.3$ and $1$, we find that for $\gamma=0.3$, a given increase of $Q$ can accommodate more VRs in the game than $\gamma=1$.

Fig.~\ref{fig:profit_vs_Q} shows both $S^{NSP}$ and $S^{GLB}$ versus $Q$ for the two schemes for a given $\gamma=1$. We can see that the revenues of both schemes increase with the growth of $Q$. It is shown that the profit $S^{NSP}$ gained by the NUPS is higher than the one gained by the UPS, while the UPS outperforms the NUPS in terms of both $S^{GLB}$ and $S^{BH}$.
\subsection{Individual VR Performance}
In this subsection, we investigate the performance of each individual VR, including the price charged to them for renting an SBS per month, and the fractions of the SBSs they rent from the NSP. We fix $\gamma=0.5$ and choose a large storage size of $Q=500$ for ensuring that all the VRs can be included. Fig.~\ref{fig:individual_price} shows the price charged to each VR for renting an SBS. The VRs are arranged according to their popularity order, ranging from $\mathcal V_1$ to $\mathcal V_{15}$, with $\mathcal V_1$ having the highest popularity and $\mathcal V_{15}$ the lowest one. We can see from the figure that in the NUPS, the price for renting an SBS is higher for the VRs having a higher popularity than those with a lower popularity. By contrast, in the UPS, this price is fixed for all the VRs. Fig.~\ref{fig:individual_fraction} shows the specific fraction of the rented SBSs at each VR. In both schemes, the VRs associated with a high popularity tend to rent more SBSs. The UPS in fact represents an instance of the water-filling algorithm. Furthermore, the UPS seems more aggressive than the NUPS, since the less popular VRs of the UPS are more difficult to rent an SBS, and thus these VRs are likely to be excluded from the game with a higher probability.
\section{Conclusions}\label{sec:conclusion}
In this paper, we considered a commercial small-cell caching system consisting of an NSP and multiple VRs, where the NSP leases its SBSs to the VRs for gaining profits and for reducing the costs of back-haul channel transmissions, while the VRs, after storing popular videos to the rented SBSs, can provide faster transmissions to the MUs, hence gaining more profits. We proposed a Stackelberg game theoretic framework by viewing the SBSs as a type of resources. We first modeled the MUs and SBSs using two independent PPPs with the aid of stochastic geometry, and developed the probability expression of direct downloading. Then, based on the probability derived, we formulated a Stackelberg game for maximizing the average profit of the NSP as well as individual VRs. Next, we investigate the Stackelberg equilibrium by solving the associated non-convex optimization problem. We considered a non-uniform pricing scheme and an uniform pricing scheme. In the former scheme, the prices charged to each VR for renting an SBS are different, while the latter imposes the same price for each VR. We proved that the non-uniform pricing scheme can effectively maximize the profit of the NSP, while the uniform one maximizes the sum profit of the NSP and the VRs. Furthermore, we derived a relationship between the optimal pricing of renting an SBS, the fraction of SBSs rented by each VR, the storage size of each SBS and the popularity of the VRs. We verified by Monte-Carlo simulations that the direct downloading probability under our PPP model is consistent with our derived results. Then we provided several numerical results for showing that the proposed schemes are effective in both pricing and SBSs allocation.
\appendices
\section{Proof of Theorem~\ref{the:Prob_cov}}\label{app:prof_theo_1}
Recall that the SBSs allocated to the VR $\mathcal V_v$ and cache $\mathcal G_f$ are modeled as a ``thinned'' HPPP $\Phi_{v,f}$ having the intensity of $\frac1 F\tau_{v}\lambda$. We consider a typical MU $\mathcal M$ who wishes to connect to the nearest SBS $\mathcal B$ in $\Phi_{v,f}$. The event $\mathcal E_{v,f}$ represents that this SBS can support $\mathcal M$ with an SINR no lower than $\delta$, and thus $\mathcal M$ can obtain the desired file from the cache of $\mathcal B$.

We carry out the analysis on $\Pr(\mathcal E_{v,f})$ for the typical MU $\mathcal M$ located at the origin. Since the network is interference dominant, we neglect the noise in the following. We denote by $z$ the distance between $\mathcal M$ and $\mathcal B$, by $x_Z$ the location of $\mathcal B$, and by $\rho(x_Z)$ the received SINR at $\mathcal M$ from $\mathcal B$. Then the average probability that $\mathcal M$ can download the desired video from $\mathcal B$ is
\begin{align}\label{equ:probability_femto}
&\Pr(\rho(x_Z)\ge\delta)\nonumber\\&=\int_{0}^{\infty}\Pr\left(\left.\frac{h_{x_Z}z^{-\alpha}}{\underset{x\in\Phi\backslash\{x_Z\}}{\sum}h_{x}\left\Vert x\right\Vert ^{-\alpha}}\ge\delta\right| z\right)f_{Z}\left(z\right)\text{d}z \nonumber\\
&=\int_{0}^{\infty}\Pr\left(\left.h_{x_Z}\ge\frac{\delta\left(\underset{x\in\Phi\backslash\{x_Z\}}{\sum}h_{x}\left\Vert x\right\Vert ^{-\alpha}\right)}{z^{-\alpha}}\right|z\right)\nonumber\\&\qquad\qquad\qquad 2\pi \frac1 F\tau_{v}\lambda z \exp\left(-\pi \frac1 F\tau_{v}\lambda z^2\right)\:\text{d}z \\
&=\int_{0}^{\infty}\mathbb{E}_{I}\left(\exp\left(-z^{\alpha}\delta I\right)\right)2\pi \frac1 F\tau_{v}\lambda z\exp\left(-\pi \frac1 F\tau_{v}\lambda z^2\right)\:\text{d}z\nonumber,
\end{align}

where we have $I\triangleq\underset{x\in\Phi\backslash\{x_Z\}}{\sum}h_{x}\left\Vert x\right\Vert ^{-\alpha}$, and the PDF of $z$, i.e., $f_{Z}\left(z\right)$, is derived by the null probability of the HPPP $\Phi_{v,f}$ with the intensity of $\frac1 F\tau_{v}\lambda$. More specifically in $\Phi_{v,f}$, since the number of the SBSs $k$ in an area of $A$ follows the Poisson distribution, the probability of the event that there is no SBS in the area with the radius of $z$ can be calculated as~\cite{Stoyan:book}
\begin{equation}\label{eq:void_pro_11}
\Pr(k=0\mid A=\pi z^2)=e^{-A \frac1 F\tau_{v}\lambda}\frac{(A \frac1 F\tau_{v}\lambda)^k}{k!}=e^{-\pi z^2 \frac1 F\tau_{v}\lambda}.
\end{equation}
By using the above expression, we arrive at $f_Z(z)=2\pi \frac1 F\tau_{v}\lambda z\exp\left(-\pi \frac1 F\tau_{v}\lambda z^2\right)$. Note that the interference $I$ consists of $I_1$ and $I_2$, where $I_1$ emanates from the SBSs in $\Phi$ excluding $\Phi_{v,f}$, while $I_2$ is from the SBSs in $\Phi_{v,f}$ excluding $\mathcal B$. The SBSs contributing to $I_1$, denoted by $\Phi_{\overline{{v,f}}}$, have the intensity of $\left(1-\frac1 F\tau_{v}\right)\lambda$, while those contributing to $I_2$ have the intensity of $\frac1 F\tau_{v}\lambda$.

Correspondingly, the calculation of $\mathbb{E}_{I}\left(\exp\left(-z^{\alpha}\delta I\right)\right)$ will be split into the product of two expectations over $I_1$ and $I_2$. The expectation over $I_1$ is calculated as
\begin{align}\label{equ:expect_I_1}
&\mathbb{E}_{I_1}\left(\exp\left(-z^{\alpha}\delta I_1\right)\right)\nonumber\\&\overset{\left(a\right)}{=} \mathbb{E}_{\Phi_{\overline{{v,f}}}}\left(\underset{x\in\Phi_{\overline{{v}}}}{\prod} \int_{0}^{\infty}\exp\left(-z^{\alpha}\delta h_{x}\left\Vert x\right\Vert ^{-\alpha}\right)\exp(-h_{x})\text{d}h_{x}\right)   \nonumber \\
&\overset{\left(b\right)}{=}\exp\left(-\left(1-\frac1 F\tau_{v}\right)\lambda\int_{\mathbb{R}^{2}}\left(1-\frac{1}{1+z^{\alpha}\delta\left\Vert x_{k}\right\Vert ^{-\alpha}}\right)\text{d}x_{k}\right) \nonumber \\
&=\exp\left( -2\pi\left(1-\frac1 F\tau_{v}\right)\lambda\frac{1}{\alpha}z^2\delta^{\frac{2}{\alpha}}B\left(\frac{2}{\alpha},1-\frac{2}{\alpha}\right)\right),\nonumber \\
&=\exp\left(-\pi\left(1-\frac1 F\tau_{v}\right)\lambda C(\delta,\alpha)z^2\right),
\end{align}
where $(a)$ is based on the independence of channel fading, while $(b)$ follows from $\mathbb{E}\left(\underset{x}{\prod}u\left(x\right)\right)=\exp\left(-\lambda\int_{\mathbb{R}^{2}}\left(1-u\left(x\right)\right)\text{d}x\right)$, where $x\in \Phi$ and $\Phi$ is an PPP in $\mathbb{R}^{2}$ with the intensity $\lambda$ \cite{Daley:book}, and $C(\delta,\alpha)$ has been defined as $\frac{2}{\alpha}\delta^{\frac{2}{\alpha}}B\left(\frac{2}{\alpha},1-\frac{2}{\alpha}\right)$.

The expectation over $I_2$ has to take into account $z$ as the distance from the nearest interfering SBS. Then we have
\begin{align}\label{equ:expect_I_2}
&\mathbb{E}_{I_2}\left(\exp(-z^{\alpha}\delta I_2)\right)\nonumber\\&=\exp\left(-\frac1 F\tau_{v}\lambda 2\pi\int_{z}^{\infty}\left(1-\frac{1}{1+ z^{\alpha}\delta r^{-\alpha}}\right)r\text{d}r\right)\nonumber\\
&\overset{(a)}{=}\exp\left(-\frac1 F\tau_{v}\lambda \pi\delta^{\frac{2}{\alpha}}z^{2}\frac{2}{\alpha}\int_{\delta^{-1}}^{\infty}\frac{\kappa^{\frac{2}{\alpha}-1}}{1+\kappa}\;\text{d}x\right) \\
&\overset{(b)}{=}\exp\left(-\frac1 F\tau_{v}\lambda\pi\delta z^{2}\frac{2}{\alpha-2}\; {{}_2}F_{1}\left(1,1-\frac{2}{\alpha};2-\frac{2}{\alpha};-\delta\right)\right)\nonumber,
\end{align}
where $\left(a\right)$ defines $\kappa\triangleq\delta^{-1}z^{-\alpha}r^{\alpha}$, and ${{}_2}F_{1}(\cdot)$ in $(b)$ is the hypergeometric function. As we defined ${A}(\delta,\alpha)=\frac{2\delta}{\alpha-2}\; {{}_2}F_{1}\left(1,1-\frac{2}{\alpha};2-\frac{2}{\alpha};-\delta\right)$, by substituting (\ref{equ:expect_I_1}) and (\ref{equ:expect_I_2}) into (\ref{equ:probability_femto}), we have
\begin{align}\label{equ:Pr_xZ}
&\Pr(\rho(x_Z)\ge\delta)=\int_{0}^{\infty}\exp\left(-\pi \left(1-\frac1 F\tau_{v}\right)\lambda C(\delta,\alpha)z^2\right)\nonumber\\&\exp\left(-\pi \frac1 F\tau_{v}\lambda z^{2}\emph{A}(\delta,\alpha)\right)2\pi \frac1 F\tau_{v}\lambda z\exp\left(-\pi \frac1 F\tau_{v}\lambda  z^{2}\right)\text{d}z \nonumber\\&=\frac{\frac1 F\tau_{v}}{C(\delta, \alpha)(1-\frac1 F\tau_{v})+{A}(\delta,\alpha)\frac1 F\tau_{v}+\frac1 F\tau_{v}}.
\end{align}
This completes the proof.
\hfill$\blacksquare$
\section{Proof of Lemma~\ref{lem:Prob_cov_xi_one}}\label{app:prof_lem_2}
By applying Lagrangian multipliers to the objective function, we have
\begin{multline}\label{equ:lem_Lag}
L(\textbf{s},\mu,\boldsymbol{\nu})=\\\sum_{j=1}^V s_j+\mu\left(\sum_{j=1}^V\sqrt{\frac{\Gamma_j}{s_j}}-(V\Lambda+\Theta)\sqrt{\frac{\lambda}{\Lambda s^{bh}}}\right)-\sum_{j=1}^V \nu_js_j,
\end{multline}
where $\mu$ and $\nu_j$ are non-negative multipliers associated with the constraints $\sum_{j=1}^V\sqrt{\frac{\Gamma_j}{s_j}}-(V\Lambda+\Theta)\sqrt{\frac{\lambda}{\Lambda s^{bh}}}\le 0$ and $s_j\ge 0$, respectively. Then the KKT conditions can be written as
\begin{equation}\label{equ:kkt_lem}
\begin{split}
\frac{\partial L(\textbf{s},\mu,\boldsymbol{\nu})}{\partial s_j}&=0,~\forall j=1,\cdots,V,\\
\mu\left(\sum_{j=1}^V\sqrt{\frac{\Gamma_j}{s_j}}-(V\Lambda+\Theta)\sqrt{\frac{\lambda}{\Lambda s^{bh}}}\right)&=0,~\text{and}~\nu_j s_j=0,~\forall j.
\end{split}
\end{equation}

From the first line of Eq. (\ref{equ:kkt_lem}), we have
\begin{equation}\label{equ:Lag_result}
s_j=\sqrt[3]{\frac{\mu^2 \Gamma_j}{4(1-\nu_j)^2}}.
\end{equation}
Obviously, we have $s_j\neq0$, $\forall j$, otherwise the constraint $\sum_{j=1}^V\sqrt{\frac{\Gamma_j}{s_j}}-(V\Lambda+\Theta)\sqrt{\frac{\lambda}{\Lambda s^{bh}}}\le 0$ cannot be satisfied. Thus, we have $\nu_j=0$, $\forall j$. Furthermore, we have $\mu\neq0$ according to Eq.~(\ref{equ:Lag_result}) since $s_j$ is non-zero. This means that $\sum_{j=1}^V\sqrt{\frac{\Gamma_j}{s_j}}-(V\Lambda+\Theta)\sqrt{\frac{\lambda}{\Lambda s^{bh}}}= 0$.

By substituting Eq.~(\ref{equ:Lag_result}) into this constraint, we have
\begin{equation}\label{equ:lem_mu}
\sqrt[3]{\mu}=\frac{\sqrt{\Lambda s^{bh}}\sum_{j=1}^V\sqrt[3]{2\Gamma_j}}{{\sqrt{\lambda}(V\Lambda+\Theta)}}.
\end{equation}
Then it follows that
\begin{equation}\label{equ:lem_s_j}
s_j=\frac{\Lambda s^{bh}\left(\sum_{v=1}^V\sqrt[3]{\Gamma_v}\right)^2\sqrt[3]{\Gamma_j}}{{\lambda(V\Lambda+\Theta)^2}}.
\end{equation}
This completes the proof.
\hfill$\blacksquare$
\section{Proof of Theorem~\ref{the:optimal_all_one_speical}}\label{app:prof_theo_2}
As discussed in Eq.~(\ref{equ:condition}) and Eq.~(\ref{equ:F_constraint}), we have proved that $Q>\frac{NC(\delta,\alpha)\left(\sum_{j=1}^V\sqrt[3]{\frac{q_j}{q_V}}-V\right)}{A(\delta,\alpha)-C(\delta,\alpha)+1}$ is a sufficient condition for the optimal solution in Eq.~(\ref{equ:lem_s_v}). In other words, as long as $Q$ is satisfied, we have the conclusion that the solution in Eq.~(\ref{equ:lem_s_v}) is optimal and $\xi_v=1$, $\forall v$.

Next, we prove the necessary aspect. Without loss of generality, we assume that
\begin{multline}\label{equ:Q_not_satisfy}
\frac{NC(\delta,\alpha)\left(\sum_{j=1}^{V-1}\sqrt[3]{\frac{q_j}{q_{V-1}}}-V+1\right)}{A(\delta,\alpha)-C(\delta,\alpha)+1}<Q\le\\\frac{NC(\delta,\alpha)\left(\sum_{j=1}^V\sqrt[3]{\frac{q_j}{q_V}}-V\right)}{A(\delta,\alpha)-C(\delta,\alpha)+1}.
\end{multline}
This leads to $s_V\ge \frac{\Gamma_vs^{bh}}{\Lambda\lambda}$, and the VR $\mathcal V_V$ will be excluded from the game. In this case, we have $\xi_j=1$, $j=1,\cdots,V-1$, and \emph{Problem~\ref{prb:nsp_pro_rw_rw}} will be rewritten as follows.
\begin{problem}\label{prb:nsp_pro_rw_rw_appendix}
We rewrite \emph{Problem \ref{prb:nsp_pro_rw_rw}} as
\begin{equation}\label{equ:NSP_optimization_rw_appendix}
\begin{split}
\min_{\textbf{s}\succeq\textbf{0}}~&\sum_{j=1}^{V-1} s_j,\\
\text{s.t.}~&\sum_{{j=1}}^{V-1} \sqrt{\frac{\Gamma_j}{s_j}}\le ((V-1){\Lambda}+ \Theta)\sqrt{\frac{\lambda}{\Lambda s^{bh}}}.
\end{split}
\end{equation}
\end{problem}
Similar to the proof of \emph{Lemma~\ref{lem:Prob_cov_xi_one}}, and combined with the constraint of $Q$ in Eq.~(\ref{equ:Q_not_satisfy}), the optimal solution of \emph{Problem~\ref{prb:nsp_pro_rw_rw_appendix}} is given by
\begin{equation}\label{equ:lem_s_v_appendix}
\hat s_v=\begin{cases}\frac{\Lambda s^{bh}\left(\sum_{j=1}^{V-1}\sqrt[3]{\Gamma_j}\right)^2\sqrt[3]{\Gamma_v}}{{\lambda((V-1)\Lambda+\Theta)^2}}, &\quad v=1,\cdots,V-1,\\\qquad\qquad \infty,&\quad v=V.
\end{cases}
\end{equation}
We can see that the optimal solution given in Eq.~(\ref{equ:lem_s_v_appendix}) contradicts to the optimal solution of \emph{Problem \ref{prb:nsp_pro_rw_rw}} given in Eq.~(\ref{equ:lem_s_v}). Hence, $Q>\frac{NC(\delta,\alpha)\left(\sum_{j=1}^V\sqrt[3]{\frac{q_j}{q_V}}-V\right)}{A(\delta,\alpha)-C(\delta,\alpha)+1}$ is a necessary condition for finding the optimal solution in Eq.~(\ref{equ:lem_s_v}). This completes the proof.\hfill$\blacksquare$
\section{Proof of Lemma~\ref{lem:U_compare}}\label{app:prof_lem_3}
Consider $v_1,v_2=1,\cdots,V$ and $v_1=v_2+1$. Then we prove that $U_{v_1}>U_{v_2}$. We have
\begin{equation}\label{equ:lemma_3_appendix}
\begin{split}
&U_{v_1}=\frac{NC(\delta,\alpha)\left(\sum_{j=1}^{v_1}\sqrt[3]{\frac{q_j}{q_{v_1}}}-v_1\right)}{A(\delta,\alpha)-C(\delta,\alpha)+1}=\\&
\frac{NC(\delta,\alpha)\left(\sum_{j=1}^{v_2}\sqrt[3]{\frac{q_j}{q_{v_1}}}-v_2+\sum_{j=v_2+1}^{v_1}\sqrt[3]{\frac{q_j}{q_{v_1}}}-(v_1-v_2)\right)}{A(\delta,\alpha)-C(\delta,\alpha)+1}\\&
=\frac{NC(\delta,\alpha)\left(\sum_{j=1}^{v_2}\sqrt[3]{\frac{q_j}{q_{v_1}}}-v_2\right)}{A(\delta,\alpha)-C(\delta,\alpha)+1}\\&\overset{(a)}{>}
\frac{NC(\delta,\alpha)\left(\sum_{j=1}^{v_2}\sqrt[3]{\frac{q_j}{q_{v_2}}}-v_2\right)}{A(\delta,\alpha)-C(\delta,\alpha)+1}=U_{v_2},
\end{split}
\end{equation}
where $(a)$ comes from the fact that $q_{v_1}<q_{v_2}$. This completes the proof.\hfill$\blacksquare$
\section{Proof of Lemma~\ref{lem:U_compare_xi_result}}\label{app:prof_lem_4}
It is plausible that if $\mathcal L$ can only keep at most $v$ VRs, it has to retain the $v$ most popular VRs to maximize its profit. Let us now prove that if $\mathcal L$ keeps $(v+w)$ VRs, $w=1,\cdots,V-v$, in the game, it cannot achieve the optimal solution for $U_{v}<Q\le U_{v+1}$.
\begin{problem}\label{prb:nsp_pro_rw_rw_appendix_rw}
In the case that $\mathcal L$ keeps $(v+w)$ VRs, we have the optimization problem of
\begin{equation}\label{equ:NSP_optimization_rw_appendix_1}
\begin{split}
\min_{\textbf{s}\succeq\textbf{0}}~&\sum_{j=1}^{v+w} s_j,\\
\text{s.t.}~&\sum_{{j=1}}^{v+w} \sqrt{\frac{\Gamma_j}{s_j}}\le ((v+w){\Lambda}+ \Theta)\sqrt{\frac{\lambda}{\Lambda s^{bh}}}.
\end{split}
\end{equation}
\end{problem}
Similar to the proof of \emph{Theorem~\ref{the:optimal_all_one_speical}}, we obtain that $Q>\frac{NC(\delta,\alpha)\left(\sum_{j=1}^{v+w}\sqrt[3]{\frac{q_j}{q_{v+w}}}-(v+w)\right)}{A(\delta,\alpha)-C(\delta,\alpha)+1}=U_{v+w}$ is the necessary condition for the $(v+w)$ VRs to participate in the game. This contradicts to the premise $U_{v}<Q\le U_{v+1}$, since we have $Q>U_{v+1}$ according to \emph{Lemma~\ref{lem:U_compare}}. Let us now consider the cases of $w'=0,-1,\cdots,1-v$. To ensure there are $(v+w')$ VRs in the game, $Q$ has to satisfy the condition that $Q>U_{v+w'}$. Since $Q>U_{v}\ge U_{v+w'}$, this implies that given $(v+w')$ VRs in the game, the NSP can achieve an optimal solution. This completes the proof.\hfill$\blacksquare$
\balance
\bibliographystyle{IEEEtran}
\bibliography{IEEEabrv,Caching}

\end{document}